\documentclass[aps,preprint,epsfig,rotate]{revtex4}
\usepackage{graphicx}
\usepackage{bm}
\usepackage{epsfig}
\input epsf
\begin{document}
%\begin{doublespace}
%
\title{Bound state properties and positron annihilation in the negatively charged Ps$^{-}$ ion.
       On thermal sources of fast positrons and annihilation $\gamma$-quanta in our Galaxy}
 \author{Alexei M. Frolov}
 \email[E--mail address: ]{alex1975frol@gmail.com}
 
%\affiliation{ITAMP, Harvard-Smithonian Center for Astrophysics, \\
%         MS 14, 60 Garden Street, Cambridge MA 02138-1516, USA}  

\affiliation{Department of Applied Mathematics, \\
 University of Western Ontario, London, Ontario N6H 5B7, Canada}

\date{\today}
%\date{April 1, 2023}

\begin{abstract}

The total energy and other bound state properties of the ground (bound) $1^{1}S$-state in the Ps$^{-}$ (or 
$e^{-}e^{+}e^{-}$) ion are determined to very high accuracy. Our best variational energy for the ground 
state in this ion equals $E$ = -0.26200507023298010777040211998 $a.u.$, which is the lowest variational 
energy ever obtained for this ion. By using our highly wave functions we evaluated (to very high accuracy) 
a number of different expectation values (or properties) of the Ps$^{-}$ ion which have never been 
determined in earlier studies. This includes a number of $\langle r^{k}_{ij} \rangle$ expectation 
values (where $5 \le k \le 11$), all independent quasi-singular Vinty-type expectation values $\langle 
\frac{{\bf r}_{ij} {\bf r}_{jk}}{r^{3}_{ij}} \rangle$, the two truly singular $\langle \frac{1}{r^{3}_{ij}} 
\rangle$ expectation values, etc. Our highly accurate expectation values of the electron-positron 
delta-function of the Ps$^{-}$ ion we have evaluated (to very high accuracy) the rates of two-, three-, 
four- and five-photon annihilation. We also discuss thermal sources of the fast positrons and annihilation 
$\gamma$-quanta located in our Galaxy \\
% DOI: 10.13140/RG.2.2.21578.03523 \\
PACS number(s): 36.10.Dr (positronium) ; 36.10.-k (exotic atoms, ions and molecules) and also 
05.30.Fk (Fermion systems and electron gas) 

\end{abstract}
\maketitle

%\newpage

\section{Introduction}

In this communication we report a number of recent results of highly accurate numerical computations which have  
been performed for the ground (bound) $1^{1}S$-state in the three-body positronium Ps$^{-}$ ion (or $e^{-} e^{+} 
e^{-}$ ion). Stability of this state in the Ps$^{-}$ ion has been shown by Hylleraas in 1947 \cite{Hyl1}. The 
ground $1^{1}S$-state is the only one bound state in the three-body positronium ion. The negatively charged 
positronium ion Ps$^{-}$ is a very interesting system for research in three-body physics, astrophysics, solid 
state physics, etc. This three-body ion includes only three leptons of equal masses and provides a large number 
of unique features. The inter-particle correlations and rules of Quantum Electrodynamics (or QED, for short) are
crucially important to describe all known properties of the Ps$^{-}$ ion, determine its life-time and allow one 
to evaluate probabilities of different decays which are possible in this ion. Formally, the positronium ion 
decays only by electron-positron annihilation into $\gamma-$rays, but it can also be involved in regular atomic 
collisions, processes and reactions, including photodetachment. In 1980's the three-body Ps$^{-}$ ion has been 
produced, detected and studied in the laboratory \cite{Mil1}, \cite{Mil2}. This stimulated extensive 
investigations of bound state properties of the Ps$^{-}$ ion. Accurate calculations of these properties have 
been performed by Bhatia and Drachman \cite{BD1}, \cite{BD2}, Ho \cite{Ho1}, \cite{Ho2} and others. In our 
earlier papers \cite{Fro2005} and \cite{Fro2015} we have also evaluated a number of bound state properties of 
this remarkable ion. 

Apparently, by analyzing the data from these old calculations we have arrived to a conclusion that it is necessary 
to recalculate some of the bound state properties. There are a number of reasons for such a conclusion. First, 
quite a few of these properties are now defined differently, e.g., they may have different signs and include some 
additional factor(s). In reality, for almost all bound state property in arbitrary three-body system we always have 
a few additional relations with other properties, and these relations arise from the basic properties of triangles 
and triangular geometry. This means that all bound state properties of an arbitrary three-body system must obey the 
general `rules of triangles'. `Traditional' definitions of some bound state properties have to be corrected in 
order to respect these rules. Second, we have to correct some inaccuracies, misprints and numerical mistakes which 
have been made in these properties in previous papers. The third reason is most important, since our current 
numerical accuracy of bound state computations significantly exceed analogous accuracy known from earlier 
calculations. Recently, by combining the ideas of analytical solutions of the Coulomb three-body problem 
\cite{Fro2001a} with highly efficient methods of numerical optimization of the non-linear parameters in trial wave 
functions \cite{Fro2001A}, we have developed the new procedure which allows one to construct extremely accurate (or 
precise, for short) solutions of the Schr\"{o}dinger equation for arbitrary, in principle, three-body systems. By 
using this procedure we have achieved a substantial progress in analytical and numerical studies of bound states 
in the Coulomb three-body systems. Now, we can determine the total energies and other bound state properties of an 
arbitrary, in principle, Coulomb three-body system to very high accuracy, or to `essentially exact' values. Highly 
accurate wave functions of the Ps$^{-}$ ion can now used to solve many long-standing problems which could not be 
determined accurately even a few years ago. In particular, in this study by using our highly accurate wave 
functions we want to re-calculate some bound state properties of the Ps$^{-}$ ion which are of great interest in 
various applications. Similar properties include various annihilation rates, a number of geometrical properties, 
lowest-order relativistic and QED corrections, photodetachment cross-sections, etc. In this study we determine and 
analyze some of these properties.  

This paper has the following structure. In the next Section we introduce three-body perimetric coordinates which 
play a central role our method to construct highly accurate variational wave functions. In Section III we define 
a number of bound state properties for the Ps$^{-}$ ion and explain our approach which is used for analytical and 
numerical computations of some fundamental three-body integrals in perimetric coordinates. Section IV includes 
important details of our calculations of various quasi-singular and singular three-body integrals. In Section V 
we determine a number of different annihilation rates which describe positron annihilation in the three-body 
Ps$^{-}$ ion. Discussion and concluding remarks can be found in the last Section. In Appendix A we discuss the 
high-temperature sources of fast positrons and annihilation $\gamma-$quanta which are known in our Galaxy. 

\section{Hamiltonian and variational wave functions}

Our main technical goal in this study is to solve the non-relativistic Schr\"{o}dinger equation $\hat{H} \Psi = E 
\Psi$ for the ground (bound) $1^{1}S$-state in the three-body Ps$^{-}$ ion. Here and everywhere below the notation 
$\hat{H}$ designates the Hamiltonian of this three-body system, $\Psi$ is the highly accurate wave function and $E 
(< 0)$ is the corresponding eigenvalue. The non-relativistic Hamiltonian of three-body Ps$^{-}$ ion is written in 
the form
\begin{eqnarray}
 \hat{H} = -\frac{\hbar^2}{2 m_{e}} \Delta_{1} - \frac{\hbar^2}{2 m_{e}} \Delta_{2} 
 -\frac{\hbar^2}{2 m_{e}} \Delta_{3} + \frac{e^2}{r_{21}} - \frac{e^2}{r_{31}} 
 - \frac{e^2}{r_{32}} \; , \; \label{Hamilt1}
\end{eqnarray}
where $\Delta_{i} = \frac{\partial^{2}}{\partial x^{2}_{i}} + \frac{\partial^{2}}{\partial y^{2}_{i}} + 
\frac{\partial^{2}}{\partial z^{2}_{i}}$ is the Laplace operator of the particle $i$, while $\pm \frac{e^2}{r_{ij}} 
= \pm \frac{e^2}{\mid {\bf r}_{i} - {\bf r}_{i} \mid}$ is the Coulomb interaction between two point particles ($i$ 
and $j$). Also in this equation $\hbar = \frac{h}{2 \pi}$ is the reduced Planck constant, which is also called the 
Dirac constant, $e$ and $- e$ are the electric charges of the positron and electron, respectively, and $m_{e}$ is 
the rest mass of the electron/positron. Below, we shall always designate the positron $e^{+}$ as the particle 3 (or 
`+'), while the two negatively charged electrons will be denoted as the particles 1 (or `-') and 2 (or `-'), 
respectively. In atomic units, where $\hbar = 1, e = 1$ and $m_e = 1$, the same Hamiltonian $\hat{H}$, 
Eq.(\ref{Hamilt1}), takes the form 
\begin{eqnarray}
 \hat{H} = -\frac{1}{2} \Delta_{1} -\frac{1}{2} \Delta_{2} - \frac{1}{2} \Delta_{3} 
  - \frac{1}{r_{32}} - \frac{1}{r_{31}} + \frac{1}{r_{21}} \; . \; \label{Hamilt}
\end{eqnarray} 
where $r_{ij} = \mid {\bf r}_i - {\bf r}_i \mid = r_{ji}$ are the three interparticle distances, or relative 
coordinates $r_{32}, r_{31}$ and $r_{21}$ which play a crucial role in this study. These three scalar coordinates 
are usedbelow as the internal coordinates. In general, in any non-relativistic three-body system there are nine 
(9 = $3 \times 3$) spatial coordinates. Three of these nine coordinates describe the motion of the center-of-mass 
of the three-body system, while three other coordinates describe orientation of this three-particle system, i.e., 
triangle of particles, in outer space. The three remaining scalar coordinates are the truly internal coordinates. 
It is very convenient to choose these internal coordinates as the three scalar interparticle distances $r_{32}, 
r_{31}$ and $r_{21}$, which are also called the relative coordinates. The exact definition of these relative 
coordinates is $r_{ij} = \mid {\bf r}_{i} - {\bf r}_{j} \mid = r_{ji}$, where ${\bf r}_{i}$ and ${\bf r}_{j}$ are 
the Cartesian coordinates of the particles $i$ and $j$. These three scalar coordinates $r_{32}, r_{31}$ and 
$r_{21}$ form the so-called natural set of internal coordinates which are translationally and rotationally 
invariant. 

These three relative coordinates are convenient to describe different interparticle interactions in three-body  
systems. The only problem with these coordinates follows from the fact that they are not truly independent of 
each other. Indeed, for these three scalar coordinates the following triangle conditions (or constraints) are 
always hold: $\mid r_{ik} - r_{jk} \mid \le r_{ij} \le r_{ik} + r_{jk}$, where $(i,j,k) = (1,2,3)$. These 
constraints complicate (often significantly) the two crucially important steps of any highly accurate method: 
(1) derivation of analytical formulas for many three-body integrals, and (2) numerical optimization of the 
non-linear parameters in variational three-body expansions. Therefore, the relative coordinates cannot be 
considered as optimal internal coordinates for three-body systems. However, there are three scalar perimetric 
coordinates $u_1, u_2, u_3$ which are simply (even linearly) related to the relative coordinates and vice 
versa:  
\begin{eqnarray}
  & & u_1 = \frac12 (r_{21} + r_{31} - r_{32}) \; \; \; , \; \; \; r_{32} = u_2 + u_3 \; \; 
  , \; \nonumber \\
  & & u_2 = \frac12 (r_{21} + r_{32} - r_{31}) \; \; \; , \; \; \; r_{31} = u_1 + u_3 \; \; 
  , \; \label{ucoord} \\
  & & u_3 = \frac12 (r_{31} + r_{32} - r_{21}) \; \; \; , \; \; \; r_{21} = u_1 + u_2 \; \; 
  , \; \nonumber
\end{eqnarray}
where $r_{ij} = r_{ji}$ are the relative coordinates defined above. The properties of perimetric coordinates 
are unique. First, these three coordinates are independent of each other. Second, each of these three 
coordinates is always non-negative. Third, each of these coordinates varies between zero and infinity, i.e., 
$0 \le u_{i} < +\infty$. These three properties mean that the original three-dimensional space of internal 
coordinates $R_{123}$ splits into a direct product of three one-dimensional spaces, i.e., $R_{123} = U_1 
\otimes U_2 \otimes U_3$, where $u_{i} \subset U_{i}$ for $i$ = 1, 2, 3. Furthermore, in many cases the 
arising three-dimensional integrals in relative coordinates are represented as finite sums of products of 
three one-dimensional integrals in perimetric coordinates. Plus, many arising one-dimensional integrals are 
reduced to the well known analytical expressions which can be found, e.g., in \cite{GR}, \cite{AS}. Briefly,
we can say that the perimetric coordinates $u_1, u_2$ and $u_3$ are very convenient in applications to 
various three-body systems \cite{AAA}. In fact, the perimetric three-body (or triangle) coordinates were 
known to Archimedes $\approx$ 2250 years ago and Hero of Alexandria $\approx$ 2000 years ago (see, the 
article {\bf Archimed} published in \cite{BSE}). A great advantage of these coordinates for the ancient 
Greeks was obvious, since these coordinates allow one to determine the area of triangle by using only the 
lengths of its sides and ignoring any angles. In modern few-body physics they have been introduced by C.L. 
Pekeris in \cite{Pek1} (more details can be found in \cite{Beth64}, \cite{MQS}, \cite{Fro2021} and references 
therein). Note that our definition of these perimetric coordinates differs from their definition used by 
Pekeris. 
   
The perimetric coordinates $u_1, u_2$ and $u_3$ are explicitly used in some variational expansions developed 
for highly accurate, bound state calculations of three-body systems (see, e.g., \cite{Acad}, \cite{Fro2005JPB}). 
One of the most effective, flexible and accurate variational expansions for three-body systems is the exponential 
variational expansion in the perimetric coordinates. This variational expansion is also very convenient in 
applications to various three-body systems, including adiabatic (or two-center) and quasi-adiabatic systems. It 
has been applied for highly accurate computations of hundreds of bound states in many dozens three-body systems, 
including many atoms, ions, exotic systems, weakly-bound and Rydberg states, nuclear systems, etc. In this study 
all highly accurate computations of the ground $1^{1}S-$state in the Ps$^{-}$ ion are also performed with the use 
of our exponential variational expansion in the three-body perimetric coordinates. For the singlet bound states 
with $L = 0$ in two electron three-body systems this variational expansion takes the form  
\begin{eqnarray}
 \Psi_{LM} &=& \frac{1}{2} (1 + \kappa \hat{P}_{21}) \sum_{i=1}^{N} C_{i} \exp(-\alpha_{i} u_1 - \beta_{i} u_2 - 
 \gamma_{i} u_3) \nonumber \\
 &=& \frac{1}{2} \sum_{i=1}^{N} C_{i} \Bigl[ \exp(-\alpha_{i} u_1 - \beta_{i} u_2 - \gamma_{i} u_3) 
 + \exp(-\beta_{i} u_1 - \alpha_{i} u_2 - \gamma_{i} u_3) \Bigr] \; , \; \label{expu} 
\end{eqnarray}
where $u_1, u_2$ and $u_3$ are the perimetric coordinates: $u_i = \frac12 ( r_{ij} + r_{ik} - r_{jk})$, where 
$r_{ij}, r_{ik}$ and $r_{jk}$ are the three relative coordinates and $(i,j,k)$ = (1,2,3). In Eq.(\ref{expu}) 
the real numbers $\alpha_{i}, \beta_{i}, \gamma_{i}$, where $i = 1, \ldots, N$, are the non-linear parameters 
of the exponential expansions. In the exponential variational expansion Eqs.(\ref{expu}) these parameters can 
be chosen (and then optimized) as arbitrary positive, real numbers and this fact substantially simplifies 
their accurate and careful optimization (see below). The highly accurate variational wave functions are used 
to determine various bound state properties of the Ps$^{-}$ ion in its ground (bound) $1^1S(L = 0)-$state. 
Many of these properties and their combinations are of great interest in a number of applications (see below). 
These problems are considered in the next Section. 
 
\section{Bound state properties}

In general, for an arbitrary bound state in some quantum system the expectation value 
\begin{eqnarray}
  X = \frac{\langle \Psi \mid \hat{X} \mid \Psi \rangle}{\langle \Psi \mid \Psi \rangle} = 
  \langle \Psi_N \mid \hat{X} \mid \Psi_N \rangle \; , \; \label{prop1}
\end{eqnarray}
is called the bound state property $X$ of this system in a given bound state \cite{Dirac} (see also \cite{Eps}). 
This property is uniformly determined by the self-adjoint operator $\hat{X}$ and the bound state wave function 
$\Psi$. As is well known (see, e.g., \cite{Dirac}) the bound state wave functions always have the finite norms. 
Without loss of generality, in all formulas below we shall assume that the wave function $\Psi$ has the unit 
norm, i.e., in Eq.(\ref{prop1}) $\Psi = \Psi_N$. By choosing different operators $\hat{X}$ we obtain different 
expectations values, or bound states properties. In this study we deal with the Ps$^{-}$ ion which is the 
Coulomb three-body system. A number of bound state properties of this Ps$^{-}$ ion can be found in Tables I - 
IV and all of them are expressed in atomic units. Physical meaning of the bound state properties follows from 
the notation used. For instance, the expectation value $\langle r_{ij} \rangle$ in Table II means the linear 
distance between particles $i$ and $j$. Analogously, the expectation values $\langle r^{k}_{--} \rangle = 
\langle r^{k}_{21} \rangle $ and $\langle r^{k}_{+-} \rangle = \langle r^{k}_{31} \rangle$ (see Table II) 
designate the $k-$th powers of these interparticle distances. Note that in the Ps$^{-}$ ion it is convenient 
to designate particles by using the symbol `+' (positron) and `-' (electron). For instance, in this notation 
the electron-electron distance in the Ps$^{-}$ ion is $\langle r_{--} \rangle$, while the electron-positron 
distance is $\langle r_{+-} \rangle$, etc. 

The notations $\delta_{+-} = \delta({\bf r}_{+} - {\bf r}_{-}) = \delta({\bf r}_{31}), \delta_{--} = \delta({\bf 
r}_{--}) = \delta_{21}$ and $\delta_{+--} = \delta_{321}$ designate the two and three-particle delta-functions, 
respectively. The expectation values of these delta-functions $\langle \delta_{+-} \rangle, \langle \delta_{--} 
\rangle$ and $\langle \delta_{321} \rangle$ have been determined to high accuracy and our results (in atomic 
units) are shown in Table III. In this Table we also show the convergence rate (or $N-$dependence) of these  
expectation values. The expectation value of the electron-positron delta-function $\langle \delta_{+-} \rangle$ 
determines a number of multi-photon annihilation rates in the ground state of the Ps$^{-}$ ion (see below). The 
expectation value of three-body delta-function $\langle \delta_{321} \rangle$ determines the one-photon 
annihilation rate. Note also that some expectation values for the Ps$^{-}$ ion are written in the form which 
include delta-functions, e.g., $\langle \delta({\bf r}_{ij}) \hat{A} \rangle$, where $\hat{A}$ is an arbitrary 
operator written in the relative and/or perimetric coordinates. For instance, the $cusp-$value between particles 
$j$-th particles $i$ and $j$ is written in the form
\begin{eqnarray}
 \nu_{ij} = \frac{\langle \Psi \mid \delta({\bf r}_{ij}) \frac{\partial}{\partial r_{ij}} \mid \Psi 
 \rangle}{\langle \Psi \mid \delta({\bf r}_{ij}) \mid \Psi \rangle} \; \; \approx \; \; \tilde{\nu}_{ij} \; 
 = \; q_i q_j \frac{m_i m_j}{m_i + m_j}\; \; . \; \; \label{cosij}
\end{eqnarray}
For the Coulomb few-body systems this expectation value must coincide (to very good accuracy) with the predicted 
cusp value $\tilde{\nu}_{ij}$ which in atomic units always equals to the $ q_i q_j \frac{m_i m_j}{m_i + m_j}$ 
value \cite{Kato}, \cite{Shra}, \cite{Kato1}. Here $q_i$ and $q_j$ are the electric charges of the particles $i$ 
and $j$, respectively, while $m_i$ and $m_j$ are their masses. All these values must be expressed in atomic units. 
For the Ps$^{-}$ ion the expected electron-positron cusp equals -0.5, while similar electron-electron cusp equals 
0.5. In general, the numerical coincidence between the predicted and computed cusp values is a good test for the 
overall quality of variational wave functions used in calculations. 

Now, consider the basic geometrical properties of the negatively charged Ps$^{-}$ ion. First of all, we need to 
know all angles in the electron-positron triangle of particles $e^{-} - e^{+} - e^{-}$. This means we have to 
determine the expectation values of all interparticle $cosine$ functions which are designated below as $\tau_{ij}$ 
and defined by the equations:
\begin{eqnarray}
 \tau_{ij} = \langle \Psi \mid \cos \theta_{ij} \mid \Psi \rangle = \langle \cos \theta_{ij} \rangle = \langle 
 \frac{{\bf r}_{ik} \cdot {\bf r}_{jk}}{r_{ik} r_{jk}} \rangle \; \; , \; \; \label{cosij}
\end{eqnarray}
where $(i,j,k)$ = (1,2,3). For an arbitrary three-body system the sum of the three $cosine$ functions slightly 
exceed unity and equals 1 + 4 $\langle f \rangle$, where the expectation value $\langle f \rangle$ is  
\begin{eqnarray}
  \langle f \rangle = \langle \frac{u_1}{r_{32}} \frac{u_2}{r_{31}} \frac{u_3}{r_{21}} \rangle =
  2 \int_{0}^{\infty} \int_{0}^{\infty} \int_{0}^{\infty} \mid \psi(u_1, u_2, u_3) \mid^2 
  u_1 u_2 u_3 du_1 du_2 du_3 \; \; . \; \label{f} 
\end{eqnarray}
For an arbitrary three-body system the following equation holds: $\tau_{32} + \tau_{31} + \tau_{21} = 1 + 4 
\langle f \rangle$. For symmetric systems, including the Ps$^{-}$ ion, this equation takes the form $2 
\tau_{32} + \tau_{21} = 1 + 4 \langle f \rangle$. This relation is absolute, i.e., it holds for all kinds of 
variational wave functions, including precise, highly accurate, accurate, approximate, etc. 

The next group of properties includes the expectation values of two scalar products $\langle {\bf r}_{ij} \cdot 
{\bf r}_{ik} \rangle$ and $\langle {\bf p}_{i} \cdot {\bf p}_{j} \rangle$. These expectation values can be 
computed either directly, or with the use of following identities: 
\begin{eqnarray}
 \langle {\bf r}_{ij} \cdot {\bf r}_{ik} \rangle = \frac12\Bigl(\langle r^{2}_{ij} \rangle + \langle 
 r^{2}_{ik} \rangle - \langle r^{2}_{jk} \rangle\Bigr) \; \;  {\rm and} \; \; 
 \langle {\bf p}_{i} \cdot {\bf p}_{j} \rangle = \frac12\Bigl(\langle p^{2}_{i} \rangle + \langle 
 p^{2}_{j} \rangle - \langle p^{2}_{k} \rangle\Bigr) \; . \; \label{rrpp}
\end{eqnarray}
The first equation here follow from the basic triangle identity ${\bf r}_{32} + {\bf r}_{21} + {\bf r}_{13} = 
0$, which always holds for the three interparticle vectors ${\bf r}_{ij} = {\bf r}_{ik} + {\bf r}_{kj}$. The 
second identity in Eq.(\ref{rrpp}) follows from the center-of-mass equation $\Bigl( {\bf p}_{1} + {\bf p}_{2} 
+ {\bf p}_{3}\Bigr) \mid \Psi \rangle = 0$, which holds for any wave functions written in the relative and/or 
perimetric coordinates (our three internal scalar coordinates), since they are translationally and rotationally 
invariant. From Eq.(\ref{rrpp}) one can also derive the two following formulas
\begin{eqnarray}
 &&\langle f(r_{32},r_{31},r_{21}) ({\bf r}_{ij} \cdot {\bf r}_{ik})\rangle = \frac12\Bigl(\langle 
  f(r_{32},r_{31},r_{21}) r^{2}_{ij}\rangle + \langle f(r_{32},r_{31},r_{21}) 
 r^{2}_{ik}\rangle - \langle f(r_{32},r_{31},r_{21}) r^{2}_{jk}\rangle\Bigr) \nonumber \\
 &&\langle f(r_{32},r_{31},r_{21}) ({\bf p}_{i} \cdot {\bf p}_{j})\rangle = \frac12\Bigl(\langle 
 f(r_{32},r_{31},r_{21}) p^{2}_{i}\rangle + \langle f(r_{32},r_{31},r_{21}) p^{2}_{j}\rangle - 
 \langle f(r_{32},r_{31},r_{21}) p^{2}_{k}\rangle\Bigr) ,\nonumber 
\end{eqnarray}
where $f(r_{32},r_{31},r_{21})$ is an arbitrary smooth function of the three relative coordinates. More 
details about calculations of radial integral in the relative and perimetric coordinates are discussed  
below. 

\subsection{Three-body integrals in perimetric coordinates}

Let us briefly explain our approach which is extensively used in this study to determine some important 
`radial' three-body integrals in perimetric coordinates. As mentioned above three-body perimetric 
coordinates $u_1, u_2$ and $u_3$ have many advantages in applications to a large number of three- and 
few-body systems. Two main advantages follow from the facts that three perimetric coordinates $u_1, u_2$ 
and $u_3$ are independent from each other and each of them varies between 0 and $+\infty$. This explains 
why the perimetric coordinates are very convenient to determine various three-body integrals, including 
some very complex and singular integrals. To illustrate this fact let us consider the following 
three-body (or three-particle) integral 
\begin{eqnarray}
 & & {\cal F}_{k;l;n}(\alpha, \beta, \gamma) = 
 \int_0^{+\infty} \int_0^{+\infty} \int_{\mid r_{32} - r_{31} \mid}^{r_{32} + r_{31}} \exp[-\alpha 
 r_{32} - \beta r_{31} - \gamma r_{21}] r^{k}_{32} r^{l}_{31} r^{n}_{21} dr_{21} dr_{31} dr_{32} 
 \nonumber \\
 & & = 2 \int^{\infty}_{0} \int^{\infty}_{0} \int^{\infty}_{0} (u_3 + u_2)^{k} (u_3 + u_1)^{l} 
 (u_2 + u_1)^{n} \times \nonumber \\
 & & \exp[-(\alpha + \beta) u_3 - (\alpha + \gamma) u_2 - (\beta + \gamma) u_1] du_1 du_2 du_3 \; 
 \; , \; \label{mine}
\end{eqnarray}
where all indexes $k, l, n$ are integer and non-negative, while 2 is the Jacobian ($\frac{\partial r_{ij}}{\partial 
u_{k}}$) of the $r_{ij} \rightarrow u_k$ substitution. This integral is called the fundamental three-body integral, 
since the knowledge of the ${\cal F}_{k;l;n}(\alpha, \beta, \gamma)$ function allows one to determine a large number 
of different three-body integrals, which are needed to solve the original Schr\"{o}dinger equation $\hat{H} \Psi = E 
\Psi$ for a given three-body system. Applications and high efficiency of the three perimetric coordinates $u_1, u_2$ 
and $u_3$ can be demonstrated by derivation of the closed analytical formula for the integral, Eq.(\ref{e11}). Here 
we just present the final result. The explicit formula for the ${\cal F}_{k;l;n}(\alpha, \beta, \gamma)$ integral is 
written in the form
\begin{eqnarray}
 {\cal F}_{k;l;n}(\alpha, \beta, \gamma) &=& 2 \sum^{k}_{k_1=0} \sum^{l}_{l_1=0} \sum^{n}_{n_1=0} 
 C_{k}^{k_1} C_{l}^{l_1} C_{n}^{n_1} \frac{(l-l_1+k_1)!}{(\alpha + \beta)^{l-l_1+k_1+1}} 
 \frac{(k-k_1+n_1)!}{(\alpha + \gamma)^{k-k_1+n_1+1}} \times \nonumber \\
 & & \frac{(n-n_1+l_1)!}{(\beta + \gamma)^{n-n_1+l_1+1}} \; \; \; , \; \label{e11} 
\end{eqnarray}
where $C^{m}_{M}$ is the number of combinations from $M$ by $m$ (here $m$ and $M$ are integer non-negative numbers). 
The formula, Eq.(\ref{e11}), can also be written in a few different (but equivalent!) forms. For the first time this 
formula, Eq.(\ref{e11}), was derived by me in the middle of 1980's \cite{Fro1987}. As mentioned above the ${\cal 
F}_{k;l;n}(\alpha, \beta, \gamma)$ integrals play a central role in physics of three-body systems. Note that our 
three-body ${\cal F}_{k;l;n}(\alpha, \beta, \gamma)$ integral defined in Eq.(\ref{e11}) exactly coincides with 
analogous $\Gamma_{k;l;n}(\alpha, \beta, \gamma)$ integrals defined in \cite{Fro2005JPB} and in other our papers.  
However, there is an obvious difference between our ${\cal F}_{k;l;n}(\alpha, \beta, \gamma)$ integral and analogous 
integral $\Gamma_{k,l,n}(\alpha, \beta, \gamma)$ introduced in \cite{HFS} which are more appropriate for one-center 
atomic systems and not for the Ps$^{-}$ ion. The relation between these two integrals takes the form: ${\cal 
F}_{k;l;n}(\alpha, \beta, \gamma) = \Gamma_{l,k,n}(\beta, \alpha, \gamma)$. 

The second fundamental integral is a direct generalization of the integral, Eq.(\ref{e11}), and it contains 
the both relative and perimetric coordinates. This `mixed' integral is written in the form   
\begin{eqnarray}
 & &{\cal H}^{p;q;t}_{k;l;n}(\alpha, \beta, \gamma; \lambda, \mu, \nu) = 
 2 \int_0^{+\infty} \int_0^{+\infty} \int_{\mid r_{32} - r_{31} \mid}^{r_{32} + r_{31}} \exp[-\alpha r_{32} 
 - \beta r_{31} - \gamma r_{21} - \nu u_3 - \mu u_2 - \lambda u_1] \times \nonumber \\ 
 & &r^{k}_{32} r^{l}_{31} r^{n}_{21} u^{p}_{1} u^{q}_{2} u^{t}_{3} dr_{21} dr_{31} dr_{32} = 2 
 \int^{\infty}_{0} \int^{\infty}_{0} \int^{\infty}_{0} (u_3 + u_2)^{k} u^{p}_{1} (u_3 + u_1)^{l} u^{q}_{2} 
 (u_2 + u_1)^{n} u^{t}_{3} \times \nonumber \\ 
 & & \exp[-(\alpha + \beta + \nu) u_3 - (\alpha + \gamma + \mu) u_2 - (\beta + \gamma + \lambda) u_1] du_1 
 du_2 du_3 \; \; \; . \; \label{mine} 
 \end{eqnarray} 
Analytical formula for this integral takes the form 
\begin{eqnarray}
 & & {\cal H}^{p;q;t}_{k;l;n}(\alpha, \beta, \gamma; \lambda, \mu, \nu) = 2 \sum^{k}_{k_1=0} \sum^{l}_{l_1=0} 
 \sum^{n}_{n_1=0} 
 C^{k_1}_{k} C^{l_1}_{l} C^{n_1}_{n} \label{e110} \\
 & & \frac{(k + l - l_1 + t)! (k - k_1 + n_1 + q)!  (l_1 + n - n_1 + p)!}{(\alpha + \beta + \nu)^{k+l-l_1+t+1} 
 (\alpha + \gamma +  \mu)^{k-k_1+n_1+q+1} (\beta + \gamma + \lambda)^{l_1+n-n_1+p}} \nonumber 
\end{eqnarray}
and it is also clear that ${\cal H}^{0;0;0}_{k;l;n}(\alpha, \beta, \gamma; 0, 0, 0) = {\cal F}_{k;l;n}(\alpha, 
\beta, \gamma)$. Again, we have to note that our formula, Eq.(\ref{e110}), can also be written in a number 
of different (but equivalent) forms. The analytical formulas, Eqs.(\ref{e11}) and (\ref{e110}), for the two 
fundamental integrals are relatively simple and they do not lead to any numerical instabilities which can restrict 
actual computations of matrix elements of the Hamiltonian and overlap matrices. In general, these two our formulas,
Eqs(\ref{e11}) and (\ref{e110}), are very effective, and currently they are extensively used to determine the 
matrix elements and expectation values of a large number of regular properties. Based on these formulas one can 
develop a number of fast, numerically stable and relatively simple algorithms which work very well in bound state 
computations of many three-body systems. 
  
\section{Quasi-singular and singular bound state properties}

Note that all expectation values considered above are regular, i.e., they are determined for the regular operators, 
which which do not include any singular parts. However, in actual computations of bound state properties one has to 
determine a number of properties which are either quasi-singular, or true singular. Any singular property, e.g., 
$\langle r^{-k}_{ij} \rangle = \langle \frac{1}{r^{k}_{ij}} \rangle$, where $k \ge 3$ is positive integer, always 
contains the both regular and non-zero singular parts. These singular parts are represented by the corresponding 
singular operators. If some operators are represented by some sums (or differences) of a number of singular operators 
and singular parts of these operators cancel each other during summation, then we deal with the so-called 
quasi-singular operators and expectation values. Another possible reason for cancellation of singular parts in 
quasi-singular expectation values follows from the actual permutation symmetry of the wave function. If such a 
cancellation of singular parts cannot be performed, or it is incomplete, then we deal with the true singular 
three-body integrals and expectation values. Fortunately, theory of singular three-body integrals in 
relative/perimetric coordinates is relatively well developed (see, e.g., \cite{HFS} and \cite{Fro2005JPB}). 

In this study we consider the two following quasi-singular properties, or expectation values: (a) $\langle 
r^{-2}_{ij} \rangle$, where $(ij)$ = (32), (31) and (21) (or (+ -) and (- -)), and (b) $\langle \frac{{\bf 
r}_{ij} \cdot {\bf r}_{ik}}{r^{3}_{ij}} \rangle$ expectation values. First, let us consider numerical 
calculations of the $\langle r^{-2}_{ij} \rangle$ expectation values. The corresponding three-body integrals, 
which are needed in computations of all matrix elements of the $r^{-2}_{32}, r^{-2}_{31}$ and $r^{-2}_{21}$ 
operators in the exponential basis function, are written in the following general form, e.g., for the 
$r^{-2}_{32}$ operator:
\begin{eqnarray}
 {\cal F}_{-1;1;1}(a, b, c) &=& 
 \int_0^{+\infty} \int_0^{+\infty} \int_{\mid r_{32} - r_{31} \mid}^{r_{32} + r_{31}} \exp[- a r_{32} - 
 b r_{31} - c r_{21}] r^{-1}_{32} r_{31} r_{21} dr_{21} dr_{31} dr_{32} \nonumber \\
 &=& \frac{\partial^{2} {\cal F}_{-1;0;0}(a, b, c)}{\partial b \partial c} =   
 \frac{\partial^{2}}{\partial b \partial c} \Bigl\{ \frac{2 [\ln (a + b) - \ln (a + c)]}{b^{2} - c^{2}} \Bigr\} 
 \; \; . \label{r32-2}
\end{eqnarray} 
and analogous formulas for the $r^{-2}_{31}$ and $r^{-2}_{21}$ operators. The final formula for the ${\cal 
F}_{-1;1;1}(a, b, c)$ integral is 
\begin{eqnarray}
 {\cal F}_{-1;1;1}(a, b, c) = \frac{4}{(b^{2} - c^{2})^{2}} \Bigl(\frac{b}{a + c} + \frac{c}{a + b}\Bigr)
 - \frac{16 b c [\ln(a + c) - \ln(a+b)]}{(b^{2} - c^{2})^{3}} \; \; . \; \label{-1+1+1}
\end{eqnarray} 
These three-body integrals are not singular. The formula, Eq.(\ref{-1+1+1}), can directly be used in numerical 
computations only in those cases when $c$ is not close to $b$. However, if $c \rightarrow b$, then this formula 
becomes numerically unstable. In such cases, we have to introduce a small parameter $\tau = \frac{c - b}{a + b}$. 
Then, the right hand side of Eq.(\ref{-1+1+1}) is represented as a power series upon $\tau$, and it contains only 
non-negative powers of $\tau$. This power series are used in numerical calculations. In detail this method was 
described in \cite{Fro2005JPB}. There are two alternative methods which are successfully used to determine the 
three-body integral, Eq.(\ref{r32-2}), and other similar three-body integrals. The first method was well 
described in \cite{HFS}. Another (third) method, which is based on semi-perimetric coordinates, will be published 
elsewhere.  
 
The second group of expectation values includes the $\langle \frac{{\bf r}_{31} \cdot {\bf r}_{32}}{r^{3}_{31}} 
\rangle$ and $\langle \frac{{\bf r}_{31} \cdot {\bf r}_{21}}{r^{3}_{31}} \rangle$ expectation values. All these 
expectation values can be written in one of the three following general forms: $\langle 
\frac{(\cos\theta_{32})^{n_1}}{r^{2}_{32}} f_a(r_{32}, r_{31}, r_{21}) \rangle, \langle 
\frac{(\cos\theta_{31})^{n_2}}{r^{2}_{31}} f_b(r_{32}, r_{31}, r_{21}) \rangle, \langle 
\frac{(\cos\theta_{21})^{n_3}}{r^{2}_{21}} f_c(r_{32}, r_{31}, r_{21}) 
\rangle$, where $f_a(x,y,z), f_b(x,y,z), f_c(x,y,z)$ are some non-singular functions of their arguments. As 
follows from these formulas none of these expectation values is singular. Moreover, numerical calculations of 
all these matrix elements does not present any troubles, if we are working in the anti-Hylleraas coordinates 
which explicitly include one $\cos \theta_{ij}$ coordinate. In general, one can write the following relations 
for an arbitrary expectation value of the operator $\hat{A}$ in the internal three-body coordinates: 
\begin{eqnarray}
  & & \int_{0}^{\infty} \int_{0}^{\infty} \int_{-\pi}^{+\pi} \Bigl[ \Psi(r_{ij}, r_{ik}, \cos\theta_{jk}) 
  \hat{A}(r_{ij}, r_{ik}, \cos\theta_{jk}) \Psi(r_{ij}, r_{ik}, \cos\theta_{jk}) \Bigr] r^{2}_{ij} r^{2}_{ik} 
  \sin\theta_{jk} dr_{ij} dr_{ik} d\theta_{jk} \nonumber \\
  &=& \int_{0}^{\infty} \int_{0}^{\infty} \int_{\mid r_{32} - r_{31} \mid}^{r_{32} + r_{31}} \Bigl[ \Psi(r_{32}, 
  r_{31}, r_{21}) \hat{A}(r_{32}, r_{31}, r_{21}) \Psi(r_{32}, r_{31}, r_{21}) \Bigr] r_{32} r_{31} r_{21} 
  dr_{21} dr_{31} dr_{32} \nonumber \\
  &=& 2 \int^{\infty}_{0} \int^{\infty}_{0} \int^{\infty}_{0} \Bigr[ \Psi(u_{1}, u_{2}, u_{3}) \hat{A}(u_{1}, 
  u_{2}, u_{3}) \Psi(u_{1}, u_{2}, u_{3}) \Bigr] u_{1} u_{2} u_{3} du_{1} du_{2} du_{3} \; , \; \label{IntGla} 
\end{eqnarray}
where $(i,j,k) = (1,2,3)$ and $\cos\theta_{jk} = \frac{r^{2}_{ij} + r^{2}_{ik} - r^{2}_{jk}}{2 r_{ij} r_{ik}}$. 
These formulas are very useful to predict expectation values which are `potentially' singular. For instance, the 
expectation value of the $r_{32}^{-2} r_{21}^{-2} \cos\theta_{31}$ operator is not singular, but the expectation 
value of the $r_{32}^{-2} r_{31}^{-2} r_{21}^{-1} \cos\theta_{21}$ operator is singular. Quite often `singularity' 
of some expectation values simply means that the chosen (or given) internal coordinates are not appropriate to 
evaluate and investigate these particular three-body integrals and expectation values.    

In reality, in our three-body perimetric coordinates as well as in the regular Hylleraas coordinates $r_{32}, 
r_{31}, r_{21}$ analytical and/or numerical calculations of all expectation values, which include the 
$(\cos\theta_{32})^{n}$ functions, where $n \ge 2$, always generate a number of problems. Indeed, each of these 
integrals is represented as a sum of some singular integrals and we have to prove that each such a sum is a 
regular expression. The number of similar problems increases rapidly when the number $n$ grows. Unfortunately, 
such integrals are needed in a large number of applications. For example, to determine the lowest order 
relativistic corrections, which are also known as Breit's corrections, we need to calculate quite a few matrix 
elements each of which contain integrals of the form $\langle (\cos\theta_{ij})^{2} \; \; f(r_{32}, r_{31}, 
r_{21}) \rangle$.  

Another example is from the history of atomic calculations. In 1940 Vinty \cite{Vinty} proposed to determine the 
largest component of isotopic shifts in the two-electron atoms by using the $\langle \frac{{\bf r}_{31} \cdot 
{\bf r}_{32}}{r^{3}_{31}} \rangle = \langle \frac{1}{r^{2}_{31}} \cos \theta_{21} \rangle$ expectation value. 
From the very beginning his method generated a number of controversies. One of them was formulated as a statement 
that this expectation value is singular, and, therefore, Vinty just reduced an original complex problem to another 
form which makes this original problem `absolutely unsolvable'. In fact, this expectation value is not singular 
(it has been shown explicitly in our paper \cite{HFS2}). Nevertheless, some questions about this and other similar 
expectation values for other three-body systems are still remain unsolved. Here we want to finish this very long 
discussion and answer all questions which currently exist for these expectation values. First, we need to know 
how many similar integrals do exist and how many of them are truly independent for an arbitrary three-body system. 
To answer these questions let us note that the following equations: $\langle \frac{{\bf r}_{ij} \cdot {\bf 
r}_{ik}}{r^{3}_{ij}} \rangle = - \langle \frac{{\bf r}_{ij} \cdot {\bf r}_{ki}}{r^{3}_{ij}} \rangle = \langle 
\frac{{\bf r}_{ji} \cdot {\bf r}_{ki}}{r^{3}_{ij}} \rangle$ and $\langle \frac{{\bf r}_{ij} \cdot {\bf 
r}_{ij}}{r^{3}_{ij}} \rangle = \langle \frac{1}{r_{ij}} \rangle$ are always obeyed for these nine expectation 
values. Also, by multiplying the vector-operator $\frac{{\bf r}_{ij}}{r^{3}_{ij}}$ by the sum ${\bf r}_{32} + {\bf 
r}_{21} + {\bf r}_{13} = 0$ one finds three following equations for these expectation values: 
\begin{eqnarray}
  &&\langle \frac{{\bf r}_{12} \cdot {\bf r}_{32}}{r^{3}_{21}} \rangle 
  + \langle \frac{{\bf r}_{21} \cdot {\bf r}_{31}}{r^{3}_{21}} \rangle 
  = \langle \frac{1}{r_{21}} \rangle \; \; , \; \; \label{EqX1} \\
  &&\langle \frac{{\bf r}_{31} \cdot {\bf r}_{21}}{r^{3}_{31}} \rangle 
  + \langle \frac{{\bf r}_{31} \cdot {\bf r}_{32}}{r^{3}_{31}} \rangle 
  = \langle \frac{1}{r_{31}} \rangle \; \; , \; \; \label{EqX2} \\
  &&\langle \frac{{\bf r}_{32} \cdot {\bf r}_{12}}{r^{3}_{32}} \rangle 
  + \langle \frac{{\bf r}_{32} \cdot {\bf r}_{31}}{r^{3}_{32}} \rangle 
  = \langle \frac{1}{r_{32}} \rangle \; \; . \; \; \label{EqX3} 
\end{eqnarray}
As follows from these equations there are only three independent $\langle \frac{{\bf r}_{ij} \cdot {\bf 
r}_{ik}}{r^{3}_{ij}} \rangle$ expectation values in an arbitrary non-symmetric three-body systems. For 
instance, we can choose the three $\langle \frac{{\bf r}_{21} \cdot {\bf r}_{32}}{r^{3}_{21}} \rangle, 
\langle \frac{{\bf r}_{31} \cdot {\bf r}_{32}}{r^{3}_{31}} \rangle$ and $\langle \frac{{\bf r}_{32} 
\cdot {\bf r}_{21}}{r^{3}_{32}} \rangle$ expectation values as independent. For symmetric three-body 
systems only two (of three) expectation values are truly independent, since in this case Eq.(\ref{EqX2}) 
and Eq.(\ref{EqX3}) are transformed into each other when $1 \Leftrightarrow 2$.  

To finish our discussion let us produce more useful relations between our expectation values $\langle 
\frac{{\bf r}_{ij} \cdot {\bf r}_{ik}}{r^{3}_{ij}} \rangle$ and some other bound state properties. First, 
we introduce the three following operators $\hat{A}_{1} = {\bf r}_{32} \cdot {\bf p}_{1}, \hat{A}_{2} = 
{\bf r}_{31} \cdot {\bf p}_{2}$ and $\hat{A}_{3} = {\bf r}_{21} \cdot {\bf p}_{3}$. For an arbitrary 
stationary (e.g., bound) state in any three-body system we can write the following equations $\langle 
\frac{d \hat{A}_{i}}{d t} \rangle = - \imath \langle [ \hat{A}_{i}, \hat{H} ] \rangle = 0$, or $\langle 
[ \hat{A}_{i}, \hat{H} ] \rangle = 0$ \cite{Hirsch}. From here one finds the following equations for the 
expectation values 
\begin{eqnarray}
  &&\frac{1}{m_3} \langle {\bf p}_1 \cdot {\bf p}_3 \rangle - \frac{1}{m_2} \langle {\bf p}_1 \cdot 
  {\bf p}_2 \rangle = q_1 q_2 \langle   \frac{{\bf r}_{32} \cdot {\bf r}_{12}}{r^{3}_{21}} \rangle - 
  q_1 q_3 \langle \frac{{\bf r}_{31} \cdot {\bf r}_{32}}{r^{3}_{31}} \rangle \; \; , \; \; \label{EqY1} \\  
  &&\frac{1}{m_1} \langle {\bf p}_1 \cdot {\bf p}_2 \rangle - \frac{1}{m_3} \langle {\bf p}_2 \cdot 
  {\bf p}_3 \rangle = q_2 q_3 \langle \frac{{\bf r}_{32} \cdot {\bf r}_{31}}{r^{3}_{32}} \rangle - 
  q_1 q_2 \langle \frac{{\bf r}_{31} \cdot {\bf r}_{21}}{r^{3}_{21}} \rangle \; \; , \; \; \label{EqY2} \\  
  &&\frac{1}{m_2} \langle {\bf p}_2 \cdot {\bf p}_3 \rangle - \frac{1}{m_1} \langle {\bf p}_1 \cdot 
  {\bf p}_3 \rangle = q_1 q_3 \langle \frac{{\bf r}_{31} \cdot {\bf r}_{21}}{r^{3}_{31}} \rangle - 
  q_2 q_3 \langle \frac{{\bf r}_{32} \cdot {\bf r}_{12}}{r^{3}_{32}} \rangle \; \; , \; \label{EqY3} 
\end{eqnarray} 
where $m_1, m_2$ and $m_3$ are the masses of three particles, while $q_1, q_2$ and $q_3$ are their electrical 
charges. All these values must be expressed in atomic units. For the Ps$^{-}$ ion one finds $m_1 = m_2 = m_3 
= 1$ and $q_1 = q_2 = - 1$, while $q_3 = 1$. For symmetric systems ($1 \leftrightarrow 2$) the last equation, 
Eq.(\ref{EqY3}), is reduced to the identity $0 = 0$ and we have only two independent equations Eq.(\ref{EqY1}) 
and Eq.(\ref{EqY2}) in this case. It is important to emphasize the fact that equations, Eqs.(\ref{EqX1}) - 
(\ref{EqX3}), are exact, while analogous equations for the same expectation values, Eqs.(\ref{EqY1}) - 
(\ref{EqY3}), are only approximate. The actual accuracy of these equations depends upon the overall accuracy 
of the trail wave functions used.

Finally, let us discuss calculations of the expectation values $\langle \frac{1}{r^{3}_{32}} \rangle (= \langle 
\frac{1}{r^{3}_{31}} \rangle$) and $\langle \frac{1}{r^{3}_{21}} \rangle$ which are truly singular. Singularity 
of these integrals follows from the fact that the following three-body integral  
\begin{eqnarray}
 {\cal F}_{-2;1;1}(a, b, c) = 
 \int_0^{+\infty} \int_0^{+\infty} \int_{\mid r_{32} - r_{31} \mid}^{r_{32} + r_{31}} \exp[- a r_{32} - b r_{31} - 
 c r_{21}] r^{-2}_{32} r_{31} r_{21} d r_{21} d r_{31} d r_{32} \; \; \label{Fsing}
\end{eqnarray}
does not exist as a finite expression, or in other words, this integral diverges. However, we can define the 
following integral   
\begin{eqnarray}
 {\cal F}_{-2;1;1}(a, b, c; \epsilon) = 
 \int_{\epsilon}^{+\infty} \int_0^{+\infty} \int_{\mid r_{32} - r_{31} \mid}^{r_{32} + r_{31}} \exp[- a r_{32} - 
  b r_{31} - c r_{21}] r^{-2}_{32} r_{31} r_{21} d r_{21} d r_{31} d r_{32} \; \; , \; \; \label{Fsing1}
\end{eqnarray}
which is a finite non-singular integral for $\epsilon > 0$, but diverges when $\epsilon \rightarrow 0$. For this 
regular ${\cal F}_{-2;1;1}(a, b, c; \epsilon)$ integral, where $\epsilon > 0$, we can apply the following formula   
\begin{eqnarray}
 {\cal F}_{-2;1;1}(a, b, c; \epsilon) = \frac{\partial^{2} {\cal F}_{-2;0;0}(a, b, c; \epsilon)}{\partial b \; 
 \partial c} \; \; , \; \; \label{Fsing2}
\end{eqnarray}
where 
\begin{eqnarray}
 {\cal F}_{-2;0;0}(a, b, c; \epsilon) = 
 \int_{\epsilon}^{+\infty} \int_0^{+\infty} \int_{\mid r_{32} - r_{31} \mid}^{r_{32} + r_{31}} \exp[- a r_{32} - 
  b r_{31} - c r_{21}] r^{-2}_{32} dr_{21} dr_{31} dr_{32} \; \; . \; \; \label{Fsing3}
\end{eqnarray}
This ${\cal F}_{-2;0;0}(a, b, c; \epsilon)$ integral is represented as the sum of its regular $R$ and singular 
$S$ parts: ${\cal F}_{-2;0;0}(a, b, c; \epsilon) = R_{-2;0;0}(a, b, c) + S_{-2;0;0}(a, b, c; \epsilon)$, where 
the regular part is 
\begin{eqnarray}
 R_{-2;0;0}(a, b, c) &=& 2 \; \frac{(a + c) \ln (a + c) - (a + b) \ln (a + b) )}{( b^{2} - c^{2} )} + 
 \frac{2}{b + c} \nonumber \\ 
 &=& R^{\ln}_{-2;0;0}(a, b, c) + R^{f}_{-2;0;0}(a, b, c) \; . \; \; \label{Fsing4}
\end{eqnarray}
The second term in the right side of this equation $R^{f}_{-2;0;0}(a, b, c) = \frac{2}{b + c}$ is called the final 
term (or final contribution), while the first term is called the logarithmic term which is designated below as 
$R^{\ln}_{-2;0;0}(a, b, c)$. The singular part of the integral, Eq.(\ref{Fsing2}) takes the form 
\begin{eqnarray}
 S_{-2;0;0}(a, b, c; \epsilon) = \frac{2}{b + c} (\psi(1) - \ln \epsilon) = - \frac{2}{b + c} (\gamma_E + 
 \ln \epsilon) \; \; , \; \; \label{Fsing45}
\end{eqnarray} 
where $\psi(x)$ is the $digamma$-function \cite{AS}. For positive integer $n$ we have $\psi(n) = -\gamma_E + 
\sum^{n-1}_{m=1} \frac{1}{m}$, where $\gamma_E = - \psi(1) \approx 0.5772156649\ldots$ is the Euler constant. 

Now by using all these formulas one can say that the limit (at $\epsilon \rightarrow 0$) of the integral ${\cal 
F}_{-2;0;0}(a, b, c; \epsilon)$, Eq.(\ref{Fsing3}), does not exist as a final expression, since its singular part, 
$S_{-2;0;0}(a, b, c; \epsilon)$ becomes infinite when $\epsilon$ approaches zero. However, the difference of this 
integral ${\cal F}_{-2;0;0}(a, b, c; \epsilon)$ and its singular part $S_{-2;0;0}(a, b, c; \epsilon)$ is a well 
defined expression which always finite, does not depend upon $\epsilon$ and equals to the regular part 
$R_{-2;0;0}(a, b, c)$ defined in Eq.(\ref{Fsing4}). In other words, the numerical value of the regularized 
integral, Eq.(\ref{Fsing3}), equals to this $R_{-2;0;0}(a, b, c)$ value which is regular and always finite. In 
our calculations we need to determine the ${\cal F}_{-2;1;1}(a, b, c; \epsilon)$ integral where $\epsilon$ is very 
small, positive and finite. The integral Eq.(\ref{Fsing1}) is the second-order derivative of the integral 
Eq.(\ref{Fsing2}). This leads to the following formulas
\begin{eqnarray}
 R_{-2;1;1}(a, b, c) = \frac{\partial^{2} R^{\ln}_{-2;0;0}(a, b, c)}{\partial b \; \partial c} + \frac{4}{(b + 
 c)^{3}} \; ; \; S_{-2;1;1}(a, b, c; \epsilon) = - \frac{4}{(b + c)^{3}} (\gamma_E + \ln \epsilon) , 
 \label{Fsing5}
\end{eqnarray}
where the second-order partial derivative in the right-hand side of the last equation is written in the form 
\begin{eqnarray}
 \frac{\partial^{2} R^{\ln}_{-2;0;0}(a, b, c)}{\partial b \; \partial c} &=& 16 \frac{[(a + c) \ln (a + c) 
 - (a + b) \ln (a + b)] b c}{(c^{2} - b^{2})^{3}} \nonumber \\
 &-& 4 \frac{[\ln (a + c) + 1] b}{(c^{2} - b^{2})^{2}} - 4 \frac{[\ln (a + b) + 1] 
 c}{(c^{2} - b^{2})^{2}} \; \; . \; \; \label{Fsing51}
\end{eqnarray} 

The same logic, which we have applied above to the ${\cal F}_{-2;0;0}(a, b, c; \epsilon)$ integral, also works 
for the ${\cal F}_{-2;1;1}(a, b, c; \epsilon)$ integral. Indeed, the limit of the ${\cal F}_{-2;1;1}(a, b, c; 
\epsilon)$ integral when $\epsilon \rightarrow 0$ simply does not exist, but this limit does exist for the 
difference of this integral and its singular part, i.e., for the ${\cal F}_{-2;1;1}(a, b, c; \epsilon) - 
S_{-2;1;1}(a, b, c; \epsilon)$ value. In fact, such a limit is finite and its is a regular function of the three 
parameters $a, b$ and $c$. As follows from our formulas presented above this limit equals to the $R_{-2;1;1}(a, b, 
c)$ value defined in he first equation from Eq.(\ref{Fsing5}). This approach also works perfectly in the general 
case, i.e., for arbitrary three-body integrals of the form ${\cal F}_{-k;m;n}(a, b, c; \epsilon)$ which contain 
singularities and diverge when $\epsilon \rightarrow 0$. Note also that for these three-body integrals we have 
predicted and found \cite{HFS} all singularities, which can be classified as follows: the lowest-order singularity 
which is proportional to the factor $\simeq (\psi(1) - \ln \epsilon) = - (\gamma_E + \ln \epsilon)$, the first- and 
higher-order singularities which are proportional to the factors $\frac{1}{\epsilon^{p}}$, where $p = 1, 2, 3, 
\ldots$. Besides these singularities, there are no other singularities for the three-particle ${\cal F}_{-k;m;n}(a, 
b, c; \epsilon)$ integrals. The same statement is true for the different three-particle integrals such as 
${\cal F}_{-p;-1;n}(a, b, c; \epsilon)$ \cite{HFS}. Very likely, a similar conclusion is also true for the three-body 
${\cal F}_{-p;-q;n}(a, b, c; \epsilon)$ integrals, where $min (p, q) \ge 1$. For such integrals one finds an 
additional complication which follows from the fact that to investigate singularities we need to apply the two 
infinitesimally small parameters $\epsilon_1$ and $\epsilon_2$ (not one $\epsilon$ which we have used in our 
analysis). Analogous singular three-body ${\cal F}_{-p;-q;-s}(a, b, c; \epsilon_1, \epsilon_2, \epsilon_3)$ integrals 
have never been investigated in earlier studies excluding perhaps the following exponential integral, which is known 
as the Demkov's three-body integral 
\begin{eqnarray}
 {\cal F}_{-1;-1;-1}(a, b, c) = 
 \int_0^{+\infty} \int_0^{+\infty} \int_{\mid r_{32} - r_{31} \mid}^{r_{32} + r_{31}} \exp[- a r_{32} - b r_{31} - 
 c r_{21}] r^{-1}_{32} r^{-1}_{31} r^{-1}_{21} d r_{21} d r_{31} d r_{32} \; . \; \label{IntDem}
\end{eqnarray}
This form is a pure formal, since it does include any of the  infinitesimally small parameters $\epsilon_i$ ($i$ = 1, 
2, 3). This integral can be found in some applications to three-body systems, but here we cannot discuss similar 
three-body integrals in details.     

Finally, by using our formulas derived above we can write for the expectation value of the $\frac{1}{r^{3}_{32}}$ 
operator 
\begin{eqnarray}
 \langle \frac{1}{r^{3}_{32}} \rangle = \lim_{\epsilon \rightarrow 0} \Bigl[ \langle \Psi \mid \frac{1}{r^{3}_{32}} 
 \mid \Psi \rangle_{\epsilon} + 4 \pi \langle \delta({\bf r}_{32}) \rangle (\gamma_E + \ln \epsilon)\Bigr] =
 \langle \Psi \mid \frac{1}{r^{3}_{32}} \mid \Psi \rangle_{R} + 4 \pi \langle \delta({\bf r}_{32}) \rangle 
  \; \; , \; \; \label{rm332} 
\end{eqnarray}
where $\langle \Psi \mid \frac{1}{r^{3}_{32}} \mid \Psi \rangle_{R}$ is the first (or logarithmic) term in the 
regular part of this expectation value which is determined by using Eq.(\ref{Fsing51}). Here we used the following 
formula $\langle \frac{4}{(b + c)^{3}} \rangle = 4 \pi \langle \delta({\bf r}_{32}) \rangle$ which is obeyed in 
the basis of exponential functions of the relative and/or perimetric coordinates. The additional term in 
Eq.(\ref{rm332}), i.e., the $4 \pi \langle \delta({\bf r}_{32}) \rangle$ term, represents the finite contribution  
into the $\langle \frac{1}{r^{3}_{32}} \rangle$ expectation value. Therefore, we cannot simply replace the singular   
$\langle \frac{1}{r^{3}_{32}} \rangle$ expectation value by its regular part, i.e., by the $\langle 
\frac{1}{r^{3}_{32}} \rangle_{R}$ expectation value, since it produces a wrong result. Note also that similar 
finite contributions always exist (and can be found) for the expectation values of singular operators. Analogously, 
for the expectation value of the $\frac{1}{r^{3}_{21}}$ operator one finds the formula 
\begin{eqnarray}
 \langle \frac{1}{r^{3}_{21}} \rangle = \lim_{\epsilon \rightarrow 0} \Bigl[ \langle \Psi \mid \frac{1}{r^{3}_{21}} 
 \mid \Psi \rangle_{\epsilon} + 4 \pi \langle \delta({\bf r}_{21}) \rangle (\gamma_E + \ln \epsilon)\Bigr] =
 \langle \Psi \mid \frac{1}{r^{3}_{21}} \mid \Psi \rangle_{R} + 4 \pi \langle \delta({\bf r}_{21}) \rangle 
  \; \; , \; \; \label{rm332} 
\end{eqnarray}
where all expectation values in the right hand side of this equation are determined by using a few obvious interchanges 
of parameters $(a, b, c)$ and variables (1, 2, 3) in our formulas presented above for the $\langle \frac{1}{r^{3}_{21}} 
\rangle$ expectation value. For the two-electron atomic systems the last expectation value, Eq.(\ref{rm332}), is called 
the Araki-Sucher term \cite{Araki}, \cite{Sucher}. This term is an important part of the lowest order QED correction in 
the two-electron atom(s) and ions and other similar systems. Analysis and calculations of the singular $\langle 
\frac{1}{r^{4}_{32}} \rangle, \langle \frac{1}{r^{5}_{32}} \rangle, \langle \frac{1}{r^{4}_{21}} \rangle$ and $\langle 
\frac{1}{r^{5}_{21}} \rangle$ integrals and corresponding expectation values for the Ps$^{-}$ ion can be found in our 
earlier studies (see, e.g., \cite{Fro2005} and references therein). 

\section{Positron annihilation}

The most remarkable property of the negatively charged positronium Ps$^{-}$ ($= e^{-} e^{+} e^{-}$) ion is annihilation 
of the electron-positron pair which can proceed with the emission of different number of photons \cite{AB}, \cite{BLP}. 
In applications to different few-body systems, which contain positron(s), this process is also called the positron 
annihilation. The most important are the cases of two- and three-photon annihilation. The formulas for the rates of 
two-, three-, four- and five-photon annihilation of the electron-positron pair in the ground state of the Ps$^{-}$ ion 
have been derived in a number of earlier studies (for more details and references, see, e.g., \cite{Fro20101}). These 
formulas are  
\begin{eqnarray}
 && \Gamma_{2 \gamma} = 2 \; \pi \alpha^{4} c a^{-1}_{0} \Bigl[ 1 - \frac{\alpha}{\pi} \Bigl( 
  5 - \frac{\pi^{2}}{4} \Bigr) \Bigr] \; \langle \delta_{+-} \rangle \; \; sec^{-1} \; , \; \; \label{gamma2} \\
 && \Gamma_{3 \gamma} = 2 \; \frac{4 (\pi^{2} - 9)}{3 \pi} \pi \alpha^{5} c a^{-1}_{0} \langle 
 \delta({\bf r}_{+-}) \rangle = \frac{8 (\pi^{2} -  9)}{3} \alpha^{5} c a^{-1}_{0} \langle 
 \delta_{+-} \rangle \; \; sec^{-1} \; , \; \; \label{threpho} \\
  && \Gamma_{4 \gamma} = 0.274 \Bigl(\frac{\alpha}{\pi}\Bigr)^{2} \Gamma_{2 \gamma} \; \; sec^{-1} \; , \; \; \; 
  {\rm and} \; \; \; \;  \Gamma_{5 \gamma} = 0.177 \Bigl(\frac{\alpha}{\pi}\Bigr)^{2} 
  \Gamma_{3 \gamma}  \; \; sec^{-1} \; , \; \; \label{4and5pho}
\end{eqnarray} 
where $\alpha = \frac{e^{2}}{\hbar c} = 7.2973525693 \cdot 10^{-3} (\approx \frac{1}{137.04}$) is the dimensionless 
fine-structure constant, $c \approx$ 2.997294580$\cdot 10^{10}$ $cm \cdot \sec^{-1}$ is the speed of light in 
vacuum, while $a_0 \approx$ 5.291772109031$\cdot 10^{-9}$ $cm$ is the Bohr radius. In atomic units $c = 
\frac{1}{\alpha}$ and $a_0 = 1$. Our formula for the two-photon annihilation rate, Eq.(\ref{gamma2}), also 
includes the lowest order QED correction to that value \cite{HaBr}. In this paper to determine the $\Gamma_{2 
\gamma}$ value we have applied the formula, Eq.(\ref{gamma2}). The formula for the three-photon annihilation 
rate was derived in 1949 \cite{3phot} and re-derived later quite a few times in a number of papers and in some 
textbooks (see, e.g., \cite{BLP} and \cite{ItzZub}). This formula is usually derived as the non-relativistic limit 
of the general formula for the three-photon annihilation rate at arbitrary energies and momenta (see, e.g., 
\cite{Fro2007}). The both formulas in Eq.(\ref{4and5pho}) have been obtained for the Ps$^{-}$ ion from the formulas 
presented in \cite{Lepage}. The total rate $\Gamma$ of positron annihilation in the Ps$^{-}$ ion is the sum of all 
partial annihilation rates, i.e., we can write $\Gamma = \Gamma_{2 \gamma} + \Gamma_{3 \gamma} + \Gamma_{4 \gamma} 
+ \Gamma_{5 \gamma} + \Gamma_{1 \gamma} + \Gamma_{6 \gamma} + \ldots$. In reality, the total annihilation rate 
$\Gamma$ is often approximated (to very good accuracy) by the sum of two its largest components $\Gamma \approx 
\Gamma_{2 \gamma} + \Gamma_{3 \gamma}$. This leads to the following formula  
\begin{eqnarray}
 \Gamma \approx \Gamma_{2 \gamma} + \Gamma_{3 \gamma} = 2 \; \pi \alpha^{4} c a^{-1}_{0} \Bigl[ 1 - \alpha 
 \Bigl(\frac{17}{\pi} - \frac{19 \pi}{12} \Bigr) \Bigr] \; \langle \delta_{+-} \rangle \; \;  \; \; sec^{-1} 
  \; \; , \; \; \label{Gammatot}
\end{eqnarray} 
These annihilation rates determined with the use of our expectation values of electron-positron delta-functions 
can be found in Table V. 

Note that in the three-body Ps$^{-}$ ion the positron annihilation can also proceed with the emission of a single 
photon. This process is called the one-photon annihilation \cite{ChuPon}, \cite{Kru}, \cite{Fro1995}. In the 
lowest-order approximation the rate of one-photon annihilation is proportional to the expectation value of the 
triple delta-function $\langle \delta_{321} \rangle$. The exact formula for the one-photon annihilation rate in 
the Ps$^{-}$ ion is written in the form \cite{Kru}, \cite{Fro1995}:  
\begin{eqnarray}
  \Gamma_{1\gamma} = \frac{64 \pi^2}{27} \alpha^{8} c a^{-1}_{0} \langle \delta_{321} \rangle \; \; sec^{-1} 
  \; , \; \label{1phot}
\end{eqnarray} 
where the expectation value $\langle \delta_{321} \rangle$ must be taken in atomic units. In general, highly 
accurate computations of this expectation value are difficult, since it is very hard to stabilize a sufficient 
number of decimal digits in the $\langle \delta_{321} \rangle$ expectation value (see Table III). Results from 
this Table indicate clearly that even with our highly accurate wave functions we cannot stabilize even four 
decimal digits in the $\langle \delta_{321} \rangle$ expectation value. Finally, we have found that for $N$ = 
3842 the one-photon annihilation rate is $\Gamma_{1 \gamma} \approx 3.2223 \cdot 10^{-2}$ $sec^{-1}$ (see Table 
V). This numerical value of the one-photon annihilation rate $\Gamma_{1\gamma}$ is noticeably smaller than its 
value evaluated in earlier papers, e.g., $\Gamma_{1 \gamma} \approx$ 3.8249$\cdot 10^{-2}$ $sec^{-1}$ 
\cite{Fro20101}. Here we have to note that our direct formula used for numerical computations of the expectation 
value of triple delta-function is not perfect, since it often leads to relatively large oscillations of the 
computed expectation values, even for highly accurate wave functions. In the future, we want to derive some 
alternative formulas for the $\langle \delta_{321} \rangle$ expectation value.     

In strong electromagnetic fields the positron annihilation in the Ps$^{-}$ ion can also proceed as 
zero-photon annihilation, i.e., annihilation of the $(e^{+}, e^{-})$-pair when no single $\gamma-$quanta 
is emitted. Approximately, the rate of such a zero-photon annihilation $\Gamma_{0 \gamma}$ in the Ps$^{-}$ 
ion can be evaluated as $\Gamma_{0 \gamma} \approx \alpha^{2} \Gamma_{1 \gamma}$, but such a rate also 
quadratically depends upon the magnetic field ${\bf H}$ (or magnetic flux density ${\bf B}$). Zero-photon 
annihilation of the Ps$^{-}$ ions may be important in strong magnetic fields which are always exist around 
very hot, rapidly rotating O-, B- and Be-stars. The Be-stars are the usual B-stars, but they also have a 
system of rapidly rotating rings of neutral hydrogen atoms. The equatorial velocities of hydrogen atoms in 
such rings often exceed 500 $km \cdot \sec^{-1}$. 

\section{Discussion and Conclusions}

Thus, we have studied the bound state properties of the ground $1^1S(L = 0)-$state of the three-body 
Ps$^{-}$ ion which have been determined numerically to high numerical accuracy. In our calculations we 
have used highly accurate variational wave functions. The overall accuracy of our current results for 
most of the properties substantially exceeds the maximal accuracy known from previous calculations. In 
particular, the total non-relativistic energy of this (ground) bound state in the three-body ion $E$ = 
-0.26200507023298010777040211998 $a.u.$ which is the lowest variational energy ever obtained for this 
ion. This indicates clearly that the unique combination of our methods developed for analytical solution 
of the Coulomb three-body problems \cite{Fro2001a} and for highly accurate numerical computations of 
bound states in similar systems \cite{Fro2001A} works very well and provides high efficiency. We have 
also determined a large number of bound state properties of the Ps$^{-}$ ion in its ground (bound) state.
Many of these properties, e.g., all $r^{k}_{21}$ and $r^{k}_{31}$ expectation values, where $k \ge 5$, 
have never been determined (to very high accuracy) in earlier studies. 

We also investigated some singular and quasi-singular properties of the Ps$^{-}$ ion. In particular, in 
this study we have solved a `mystery' of the Vinty-type $\langle \frac{{\bf r}_{ij} \cdot {\bf 
r}_{ik}}{r^{3}_{ik}} \rangle$ expectation values. All these expectation values for the Ps$^{-}$ ion are 
now known to very good accuracy. The singular expectation values $\langle r^{-3}_{31} \rangle$ and 
$\langle r^{-3}_{21} \rangle$ have been determined to very high accuracy. Our highly accurate expectation 
values of the electron-positron delta-functions, i.e., $\langle \delta_{31} \rangle (= \langle \delta_{+-} 
\rangle)$, allowed us to determine a number of multi-photon annihilation rates in the Ps$^{-}$ ion (see, 
Table V). Numerical values of these annihilation rates are now known to very high numerical accuracy, and 
they will not be changed noticeably in future calculations. We also evaluated the numerical value of 
one-photon annihilation rate $\Gamma_{1\gamma}$ (see, Table V) and discussed a few problems which still 
remain in accurate computations of this rate. Recently, annihilation in the Ps$^{-}$ ion and other 
properties of this ion have been discussed in \cite{Dary} and \cite{Bress}. Zero-photon annihilation of 
the  Ps$^{-}$ ion is possible only in strong $EM-$fields and must be evaluated in future studies. 

As we have mentioned in the text the lowest-order relativistic and QED corrections for the ground $1^1S(L 
= 0)-$state of the three-body Ps$^{-}$ ion will be determined in our next study. Numerical calculation of 
these corrections require the closed analytical formulas for some new three-body integrals which have not 
been considered in this study. Our earlier results \cite{Fro2005PLA} (see also \cite{BD2}) for these 
values are outdated and must be corrected. Another paper will include our new formulas for the 
photodetachment cross section of the negatively charged Ps$^{-}$ ion. In general, for the Ps$^{-}$ ion one 
finds a large number of interesting problems none of which have ever been investigated. 

In this study we also investigated a close problem of thermal sources of fast positron and annihilation 
$\gamma$-quanta in our Galaxy (see, Appendix A below). In this direction we have discovered a number of 
very interesting facts about the overall intensity of such overheated sources, evaluated the numbers of 
electron-positron pairs in them (at different temperatures), etc. We have also found that it is impossible 
to see (directly) any object, e.g., gas, or atomic plasma, heated to temperatures above 350 - 400 $keV$. 
An actual observer cannot see (directly) the shape of atomic plasma heated to extremely high temperatures, 
but instead will observe only an intense flow of fast positrons, electrons and huge amount of annihilation 
$\gamma-$quanta. This phenomenon is the {\it annihilation shielding}  of overheated matter, and probably, 
it plays an important role in Astrophysics. The existence of such a high-temperature limit for photon 
optics has never been assumed either in classical, or quantum optics, where it was always believed that one 
could see all details of material objects heated to arbitrary high temperatures.  

\appendix
  \section{On thermal sources of annihilation $\gamma$-quanta in our Galaxy}
\label{A} 

Annihilation of the electron-positron pairs from the bound states of different light polyelectrons (e.g., 
the positronium Ps$^{-}$ ion, positronium Ps$_2$ quasi-molecule, etc), positron-containing light atoms, 
ions and quasi-molecules is of great interest in the both Stellar and Galactic astrophysics (see, e.g., 
\cite{DrachCJP} and references therein). In reality, all processes with positrons are still considered as 
`exotic' (or `rare') even in Stellar astrophysics. The main problem here is a short life-time of newly 
created positrons in regular stellar photospheres. However, in our Galaxy (Milky Way) there are quite a 
few known sources of fast positron and annihilation $\gamma-$quanta with energies $E_{\gamma} \approx$ 
0.5110 $MeV$. Annihilation $\gamma-$quanta and positrons from these sources propagate into the rest of 
our Galaxy in relatively large quantities. There are a number of remarkable experimental facts about 
such sources of annihilation $\gamma-$quanta and fast positrons. First of all, they have unusually high 
intensities, which lasts for many dozens and even hundreds of years (without visible weakening). Second, 
all such sources of $\gamma$-radiation are either located in the immediate vicinity proximity to the 
known black holes, or in regions where such holes are currently being formed. One of such annihilation 
sources is located almost at the center of our Galaxy, where one also finds a number of known black holes 
one of which is extremely massive (Sagittarius A$^{*}$). The evaluated mass of Sagittarius A$^{*}$ 
approximately equals 4.1 million times the mass of the Sun, while its spatial diameter has been estimated 
as 52 million kilometers. This diameter determines the so-called event horizon which is a distance from 
the center of the black hole within which nothing can escape. This black hole (Sagittarius A$^{*}$) lies 
at the center of our Milky Way Galaxy at the distance $\approx$ 26750 light years from our Solar system. 
In spatial areas located outside of event horizon, but close to this massive black hole, some components 
of the gravitational field strengths $\frac{\partial g_{\alpha\beta}}{\partial x^{\gamma}}$, or Cristoffel 
symbols $\Gamma^{\gamma}_{\alpha\beta}$, reach very large values, which are in dozens of billions times 
greater than similar values of these quantities inside of our Sun. Similar field strengths can easily 
confine even a very hot plasma which is our interest in this study.  

In general, the presence of intense sources of the fast positrons and annihilation $\gamma-$quanta, i.e., 
$\gamma-$quanta with energies $E_{\gamma} \approx$ 0.5110 $MeV$, means that there are extremely high 
temperatures in some small (local), or relatively large (extended) spatial areas. For instance, let us 
consider some relatively large $B-$, or $A-$star which has been captured by a supermassive black hole. In 
99.9\% of cases this is not a sudden event, but a process which takes quite some time, e.g., dozens of 
years. The motion and evolution of the captured star in the field of supermassive black hole is a very 
complex and interesting process.  

\vspace{5cm}

If the local equilibrium temperature $T$ in atomic plasma increases to very large values, e.g., $T \ge 150$ 
$keV$, then positrons become more and more common particles in such an overheated region. To illustrate this 
let us consider one hydrogen atom which has its `natural' volume $V_H = \frac{4 \pi}{3} a^{3}_{0}$, where 
$a_0$ is the Bohr's radius. At normal conditions such an atom contains one positively charged hydrogen 
nucleus and one electron. Let us assume that by using some boundary conditions we can somehow hold these two 
and other newly created particles in the volume $V_H$. Now, the fundamental question is: what will we see in 
this volume $V_H$ when the local temperature $T$ rises? For relatively small temperatures it is easy to 
answer this question. Indeed, until the local temperature reaches some relatively large values, e.g., $T$ = 
125 $keV$, we will not see any changes in this volume $V_H$, i.e., no new particles will arise in this heated 
volume $V_H$. However, already for $T$ = 130 $keV$ the same volume of hydrogen atom will also include 
$\approx$ 2.2 electron-positron pairs. For higher temperatures, e.g., for $T$ = 150 $keV$, in the volume 
$V_H$ we will see 9.7 $(e^{-}, e^{+})-$pairs, for $T$ = 170 $keV$ this number equals $\approx$ 31.4 $(e^{-}, 
e^{+})-$pairs and for $T$ = 200 $keV$ there are $\approx$ 126 such pairs. If the temperature 170 $keV$, then 
we can ignore the original hydrogen nucleus and atomic electron in the volume $V_H$, since their overall 
contribution into thermodynamic functions becomes negligible. Briefly, at such conditions we have to deal 
with the dense electron-positron plasma and annihilation $\gamma-$quanta. The role of incident particles, e.g., 
atomic nuclei and electrons can be ignored.   

These figures explain a significant contribution of annihilation gamma quanta and electron-positron pairs into 
the total energy release of nuclear and thermonuclear explosions. It was reported in \cite{Yamp} that up to 
11\% of all energy released in the standard 40 $kt$ explosion of a nuclear charge assembled in a multi-shell 
configuration goes into the formation of electron-positron pairs. In standard devices a significant part of 
newly created positrons will annihilate inside of the nuclear charge. In turn, annihilation $\gamma-$quanta are 
either scattered by heavy elements inside this nuclear charge (or bomb), or absorbed and then re-emitted with 
some delay and different frequencies. However, some positrons may leave the central part of the bomb and 
annihilate outside the explosion area. This produces, in particular, disturbances and interruptions in the radio 
communications. During high-temperature thermonuclear explosions, when $T \ge$ 230 $keV$, the electron-positron 
pairs are formed in significantly larger numbers and consequences of similar explosions can be catastrophic for 
any communications based on electronic devices. However, if such high temperature explosions are produced at 
very high altitudes ($h \ge 100$ $km$), or in space ($h \ge 200$ $km$), then regular tv- and radio-communications 
can be interrupted for a very long time. 

For higher temperatures the total number of newly created positrons (and electrons) increases faster with the 
temperature. For instance, for $T$ = 250 $keV$ in the same volume $V_H$ we have $N_{e^{-} e^{+}}$ = 684, for 
$T$ = 300 $keV$ our evaluation gives $N_{e^{-} e^{+}}$ = 2336 and for $T$ = 400 $keV$ this number equals 
12978. These total numbers of newly created $(e^{-}, e^{+})$-pairs have been evaluated by using the formulas 
presented below. Here we want to note that for extremely high temperatures the numbers of electron-positron 
pairs in the volume $V_H$ become extremely large, e.g., for 500 $keV$ we have $N_{e^{-} e^{+}}$ = 939305, for 
$T$ = 0.5110 $MeV$ such a number surpasses one million, while for the `fantastically' high, `nuclear' 
temperature $T$ = 5 $MeV$ the total number of electron-positron pairs created in the volume of one hydrogen 
atom $V_H$ exceeds one billion. In similar areas of stellar plasma heated to very high temperatures the 
methods of atomic physics stop working and we have to apply the methods of statistical physics \cite{LLSF}. 
Furthermore, we can neglect all incident particles, which were originally existed in these areas, and consider 
the newly created electron-positron plasma only. It is also clear that such an electron-positron plasma must 
be in thermal equilibrium with the gas of annihilation $\gamma-$quanta (or photon gas, for short). 

Thus, we can write the following `chemical' reaction: $e^{-} + e^{+} = \gamma_1 + \gamma_2 + \ldots$, which 
describes equilibrium between the hot electron-positron plasma and photon gas of annihilation $\gamma-$quanta. 
Just as in the case of chemical reactions, we write expressions for the chemical potentials of all three gases 
and obtain the following equation $\mu_{e} + \mu_{p} = \mu_{\gamma} = 0$, since the chemical potential of any 
photon gas equals zero identically. Here and everywhere below, the index $e$ means electron(s), while the 
index $p$ designates positron(s). For relatively small temperatures, e.g., for $T = 50 - 150$ $keV$, we can
 write the following, explicit formula for the chemical potentials of electron and/or positron gases (they 
both are the gases of fermions)  
\begin{eqnarray}
  \mu_{e} = T \; \ln\Bigl[ \frac{N_e}{2 V} \Bigl(\frac{2 \pi \hbar^{2}}{m T}\Bigr)^{\frac32} \Bigr] + m_e 
  c^{2} \; \; , \; \; {\rm and} \; \; \; \; 
  \mu_{p} = T \; \ln\Bigl[ \frac{N_p}{2 V} \Bigl(\frac{2 \pi \hbar^{2}}{m T}\Bigr)^{\frac32} \Bigr] + m_e 
  c^{2} \; \; , \; \; \label{chempot}
\end{eqnarray}  
respectively. In these equations $m = m_e$ is the electron/positron mass at rest, $c$ is the speed of light in 
vacuum, $\hbar$ is the reduced Plank's constant (or Dirac constant), while $N_e$ and/or $N_p$ are the numbers 
of electrons and positrons, respectively, which are located in the volume $V$. Here it is better to introduce 
the corresponding (spatial) densities of particles: $n_{e} = \frac{N_e}{V}$ and $n_{p} = \frac{N_p}{V}$. Then
from the equation of thermal equilibrium $\mu_{e} + \mu_{p} = 0$ mentioned above one finds
\begin{eqnarray}
  n_{e} n_{p} = \frac12 \Bigl(\frac{m T}{\pi \hbar^{2}}\Bigr)^{3} \; \exp\Bigl(-\frac{2 m c^{2}}{T}\Bigr) \; 
   = \frac{1}{2 \pi^{3} a^{3}_0 \alpha^{3}} T^{3} \; \exp\Bigl(-\frac{2}{T}\Bigr) \; \; , \; \label{prodnenp}
\end{eqnarray}
where the temperature $T$ is expressed in the energy units of 0.5109989500 $MeV$ and $\alpha$ is the 
dimensionless fine structure constant $\approx \frac{1}{137.04}$ (see above). Let us assume that initially, 
e.g., for $T$ = 50 $keV$  we have $N_0$ free electrons in the volume $V$, then it is easy to obtain the exact 
values of the both electron $n_{e}$ and positron $n_{p}$ densities 
\begin{eqnarray}
  n_{p} = n_{e} - n_0 = - \frac{n_{0}}{2} + \frac12 \Bigl[ n^{2}_{0} + 2 \Bigl(\frac{m T}{\pi 
  \hbar^{2}}\Bigr)^{3} \; \exp\Bigl(-\frac{2 m c^{2}}{T}\Bigr)\Bigr]^{\frac12} \; \; \label{prodnenp1a} 
\end{eqnarray}  
and  
\begin{eqnarray}  
  n_{e} = \frac{n_{0}}{2} + \frac12 \Bigl[ n^{2}_{0} + 2 \Bigl(\frac{m T}{\pi 
\hbar^{2}}\Bigr)^{3} \; \exp\Bigl(-\frac{2 m c^{2}}{T}\Bigr)\Bigr]^{\frac12} \; , \; \label{prodnenp1b} 
\end{eqnarray}
where $n_0 = \frac{N_{e}(0)}{V}$ is the spatial density of initial electrons. As follows from this formula, if 
$T \ll m c^{2}$, then the total number of newly created electron-positron pairs is exponentially small 
(negligible), but it rapidly increases with the temperature.  

At higher temperatures ($T \ge$ 170 $keV$) the formulas presented above begin to lose their accuracy. As 
mentioned above in these cases the total numbers of newly created electrons and positrons substantially exceed
the total number of initial particles, i.e., atomic electrons and atomic nuclei. Therefore, to very good 
accuracy we can assume that at these temperatures the total number of electrons in the heated area $V$ equals 
to the total number of positrons. The accuracy of such an approximation rapidly increases with the temperature
and it is already very good for $T \ge$ 170 $keV$. Therefore, their chemical potentials equal to each other 
and from the equation $\mu_{e} + \mu_{p} = 0$ mentioned above, we find $\mu_{e} = \mu_{p} = 0$. This allows us 
to produce the following formula for the total number of electrons $N_e$ and positrons $N_p$ in the heated 
volume $V$: 
\begin{eqnarray}
  N_{p} = N_{e} = \frac{(2 s + 1) V}{2 \pi^{2} {\hbar}^{3}} \; \int_{0}^{\infty} \frac{p^{2} 
  dp}{\exp\Bigl(\frac{c \sqrt{p^{2} + m^{2} c^{2}}}{T} \Bigr) + 1} = \frac{V}{\pi^{2} {\hbar}^{3}} \; 
  \int_{0}^{\infty} \frac{p^{2} dp}{\exp\Bigl(\frac{c \sqrt{p^{2} + m^{2} c^{2}}}{T} \Bigr) + 1} 
  \; \; , \; \; \label{N-eN-p1}
\end{eqnarray}  
where we have used the facts that the both electrons and positrons are fermions and spin of each of this 
particles equals $\frac12$. This equation can be re-written into a different form which is more convenient 
for the both theoretical analysis and numerical calculations. In particular, if we are dealing with the 
volume of a hydrogen atom, i.e., $V = V_H$, then the formula, Eq.(\ref{N-eN-p1}), for $N_p (= N_e)$ takes 
the form    
\begin{eqnarray} 
  N_{p} = \frac{4}{3 \pi} \alpha^{-3} \; \theta^{3} \int_{0}^{\infty} \frac{y^{2} dy}{\exp\sqrt{y^{2} + 
  \frac{1}{\theta^{2}}} + 1} \approx 1.0921766195 \cdot 10^{6} \cdot \theta^{3} \int_{0}^{\infty} 
  \frac{y^{2} dy}{\exp\sqrt{y^{2} + \frac{1}{\theta^{2}}} + 1} , \label{N-eN-p2}
\end{eqnarray}  
where $\theta = \frac{T}{m c^{2}}$ is the temperature expressed in the $m c^2$-energy unis and $N_e = N_p$. 
These formulas describe the distribution of the number(s) of positrons and/or electrons upon the temperature 
$\theta$. Again, we have to note that this formula is correct, if (and only if) the chemical potential of 
each of these two gases equals zero. Analytical computations of integrals in Eqs.(\ref{N-eN-p1}) and 
(\ref{N-eN-p2}) is not a difficult problem. Our analytical expression for the integral in Eq.(\ref{N-eN-p2}) 
is
\begin{eqnarray} 
 \int_{0}^{\infty} \frac{y^{2} dy}{\exp\sqrt{y^{2} + \frac{1}{\theta^{2}}} + 1} &=&  
  \frac{1}{\theta^{2}} \sum^{\infty}_{n=1} \frac{(-1)^{n-1}}{n} K_{2}\Bigl(\frac{n}{\theta}\Bigr) = 
 \frac{1}{\theta^{2}} \Bigl[ K_{2}\Bigl(\frac{1}{\theta}\Bigr) - \frac12 K_{2}\Bigl(\frac{2}{\theta}\Bigr)   
 + \frac13 K_{2}\Bigl(\frac{3}{\theta}\Bigr)  \nonumber \\
 &-& \frac14 K_{2}\Bigl(\frac{4}{\theta}\Bigr) + \frac15 K_{2}\Bigl(\frac{5}{\theta}\Bigr) 
 - \frac16 K_{2}\Bigl(\frac{6}{\theta}\Bigr) + \frac17 K_{2}\Bigl(\frac{7}{\theta}\Bigr) + 
 \ldots \Bigr] , \label{N-eN-p3}
\end{eqnarray}  
where $K_{2}(x)$ is the modified Bessel function of the second order. The $K_{p}(x)$ functions are also 
called the Macdoanld's functions, since H.M. Macdonald studied and introduced these functions in 1899 
\cite{Mac} (see, also discussion and references in \cite{Watson}). 

Finally, the energy of electron/positron gas takes the form 
\begin{eqnarray}
 E_e = E_p &=& \frac{V}{\pi^{2} \hbar^{3}} \int^{+\infty}_{a} \frac{c \; \sqrt{p^{2} + m^{2} c^{2}} 
 \; p^{2} dp}{\exp\sqrt{\Bigl(\frac{p c}{T}\Bigr)^{2} + \Bigl(\frac{m c^{2}}{T}\Bigr)^{2}} + 1} 
 \nonumber \\
 &=& \frac{V T}{\pi^{2}} \Bigl(\frac{T}{\hbar c}\Bigr)^{3} \Bigl\{\sum^{\infty}_{n=1} (-1)^{n-1}  
 \Bigl[ \frac{a^{3}}{2 n} K_{1}(n a) + \frac{a^{2}}{n^{2}} K_{2}(n a) + \frac{a^{3}}{2 n} K_{3}(n a) 
 \Bigr]\Bigr\} \; \; , \; \; \label{N-eN-p4}
\end{eqnarray}
where $a = \frac{m c^{2}}{T} = \frac{1}{\theta}$. Derivation of the formulas, Eqs.(\ref{N-eN-p2}) - 
(\ref{N-eN-p4}), will be published elsewhere \cite{Fro2023}. Note that the formulas Eqs.(\ref{N-eN-p2}) 
- (\ref{N-eN-p4}) are not based on any approximation and describe the properties of an arbitrary, in 
principle, Fermi gas at high and very high temperatures. The chemical potential of such a gas equals 
zero identically, e.g., in this Fermi gas of particles is in thermal equilibrium with the photon gas 
of annihilation $\gamma-$quanta. For numerical approximations and evaluations the formula, 
Eq.(\ref{N-eN-p2}), works very well, if $\theta \ge 0.15$. Formally, it has no restrictions for extremely 
high temperatures $\theta$, but our alternating series in Eqs.(\ref{N-eN-p3}) and (\ref{N-eN-p4}) can 
produce a few numerical troubles for extremely high temperatures. Detail investigation of thermodynamic 
properties of the electron-positron plasma at high and very high temperatures will be performed in our 
next study \cite{Fro2023}. 

In those cases when $T \gg m c^{2}$, e.g., for $T \approx$ 5 $MeV$, we can assume (to very good accuracy) 
that in Eq.(\ref{N-eN-p1}) $c \sqrt{p^{2} + m^{2} c^{2}} = c p$. This allows one to derive a very simple 
analytical formula (directly from Eq.(\ref{N-eN-p1})) for the total number of electrons and/or positrons 
in this volume $N_{p} = \frac{3 \zeta(3)}{2 \pi^{2} (\hbar c)^{3}} \; T^{3} V$, where $V$ is the volume 
of the heated area, $\zeta(x)$ is the Riemann function and $\zeta(3) \approx$ 1.202056903159594285399... 
\cite{zeta}. This number equals to the number of newly created electrons $N_{e}$ and to the total number 
of electron-positron pairs $N_{e^{-} e^{+}}$. As follows from this formula, the total number of arising 
positrons (and electrons) in the volume $V$ increases as the cubic function of temperature. In general, 
at these temperatures $T \ge 5$ $MeV$ the number(s) of arising electron-positron pairs are extremely large. 
For instance, the volume of a single hydrogen atom $V_H = \frac{4 \pi}{3} a^{3}_{0}$ contains (at $T$ = 5 
$MeV$) more than one billion electron-positron pairs. The corresponding energy of this gas of 
electron-positron pairs (or electron-positron gas, for short) equals $E_{e^{-} e^{+}} = \frac{7 \pi^{2}}{60 
(\hbar c)^{3}} T^{4} V$, and such an energy rapidly increases with the temperature. The total quadratic 
moment of all positrons/electrons enclosed in volume $V$ is $\langle {\bf p}^{2} \rangle = \frac{45 \; 
\zeta(5) \; \hbar^{2}}{2 \; \pi^{2}} V \Bigl(\frac{T}{\hbar c}\Bigr)^{5}$. Other properties of the 
electron-positron gas located in this volume $V$ are determined analogously. Annihilation of such an 
electron-positron gas held in relatively large confined volumes is a separate, but very interesting problem, 
which has been considered in our earlier papers \cite{Fro2007} and \cite{phot-det}. 

Analysis of thermal sources of annihilation $\gamma$-quanta in our Galaxy (and other Galaxies) indicates 
clearly that at certain thermal and gravitational conditions positrons and electrons become the two most 
common particles in some parts of the Universe. For instance, a volume (cube) of stellar matter with an 
edge of 1000 kilometers heated to a temperature of $T$ = 0.5110 $keV$ will contain approximately 
6.7483345$\cdot 10^{54}$ electron-positron pairs. Certainly, from a distance of 27,000 light years, we 
cannot see such a very small volume. However, annihilation $\gamma-$quanta and fast positrons from this 
overheated volume can be registered and observed. From this point of view, our analysis of the annihilation 
of electron-positron pairs is of considerable interest (see, e.g., \cite{Panth1}, \cite{Panth2} and 
references therein). Alternative explanations of the sources of annihilation $\gamma-$quanta are based on 
a number of unrealistic assumptions and they cannot explain extremely high intensities and relatively long 
lives of actual sources of annihilation $\gamma-$quanta which do exist in our Galaxy. 
 
From this brief discussion of the thermal sources of annihilation $\gamma-$quanta in our Galaxy, we can derive 
an important conclusion for the photon optics. This conclusion can be expressed in the following words: {\it 
due to the electromagnetic instability of the vacuum, it is impossible to see (directly) any material object 
(i.e., plasma) heated and confined at the temperatures above} 300 - 350 $keV$. In reality, instead of such a 
`static' material object, heated to extremely high temperatures, an observer will see only an intense flow of 
annihilation $\gamma-$quanta mixed with fast electrons and positrons. Briefly, this phenomenon corresponds to 
the annihilation shielding (or annihilation cooling) of any overheated material objects. The main interest 
this phenomenon presents for the overheated atomic plasma which is confined by very strong gravitational field. 
Briefly, this means that at very high temperatures the traditional optics ends and from any overheated atomic 
plasma we will always see only extremely intensive streams of outgoing annihilation $\gamma-$quanta mixed with 
large number of fast electrons and positrons. The existence of such a high-temperature limit in photon optics 
has never been assumed either in classical, or quantum optics. Moreover, in traditional optics people always 
believed that we can see all details of material `objects' heated to arbitrary high temperatures (see, e.g., 
\cite{Planck}, \cite{BornW} and references therein). Probably, the existence of high-temperature limit in 
photon optics is the most important result of our analysis of thermal sources of annihilation $\gamma$-quanta 
in our Galaxy. \\

\newpage
\newpage
\begin{table}[tbp]
   \caption{Convergence of the total energies $E$ (in $a.u.$) determined for the ground $1^{1}S-$state of 
            the Ps$^{-}$ ion. The notation $N$ is the total number of basis functions used.}
     \begin{center}
%     \scalebox{0.80}{%
     \begin{tabular}{| c | c | c |}
      \hline\hline
 $N$  & $E$ (variant $A^{(a)}$) & $E$ (variant $B$) \\ 
     \hline
 3400  & -0.26200507023298010777037030335 & -0.26200507023298010777037189975 \\ 

 3500  & -0.26200507023298010777038494512 & -0.26200507023298010777038551005 \\ 

 3600  & -0.26200507023298010777039244155 & -0.26200507023298010777039258151 \\ 

 3700  & -0.26200507023298010777039697369 & -0.26200507023298010777039684716 \\

 3800  & -0.26200507023298010777040096213 & -0.26200507023298010777040100869 \\

 3840  & -0.26200507023298010777040206466 & -0.26200507023298010777040210806 \\ 

 3842  & -0.26200507023298010777040206643 & -0.26200507023298010777040211998 \\
         \hline\hline
  \end{tabular}
  \end{center}
${}^{(a)}$Variants $A$ and $B$ represent the two different optimization strategies.  
\end{table}
\begin{table}[tbp]
   \caption{The expectation values of a number of regular properties (in atomic units) of the ground (bound) 
            $1^1S-$state in the Ps$^{-}$ ion. The notations $+$ and $-$ denote the positron and electron, 
            respectively.}
     \begin{center}
     \scalebox{0.95}{%
     \begin{tabular}{| c | c | c | c |}
      \hline\hline
  $\langle r^{-1}_{+-} \rangle$ & 0.33982102305922030648057 & $\langle r^{-1}_{--} \rangle$ & 0.15563190565248039742034 \\
  $\langle r_{+-} \rangle$ & 5.4896332523594499332956 & $\langle r_{--} \rangle$ & 
                             8.5485806550991861114230 \\
  $\langle r^{2}_{+-} \rangle$ & 48.41893722623795540990 & $\langle r^{2}_{--} \rangle$ & 
                                 93.17863384798132899897 \\ 
            \hline
  $\langle r^{3}_{+-} \rangle$ & 6.07295629623278442058$\cdot 10^{2}$ & $\langle r^{3}_{--} \rangle$ & 
                                 1.26558044787814412021$\cdot 10^{3}$ \\
  $\langle r^{4}_{+-} \rangle$ & 9.9306386797960041295$\cdot 10^{3}$ & $\langle r^{4}_{--} \rangle$ & 
                                 21.054453389258358046$\cdot 10^{4}$ \\ 
  $\langle r^{5}_{+-} \rangle$ & 2.002717783416503779$\cdot 10^{5}$ & $\langle r^{5}_{--} \rangle$ & 
                                 4.218459887428125939$\cdot 10^{5}$ \\
            \hline
  $\langle r^{6}_{+-} \rangle$ & 4.805681251065410643$\cdot 10^{6}$ & $\langle r^{6}_{--} \rangle$ & 
                                 9.999299519094784345$\cdot 10^{6}$ \\
  $\langle r^{7}_{+-} \rangle$ & 1.33846111735368123$\cdot 10^{8}$ & $\langle r^{7}_{--} \rangle$ & 
                                 2.754141872101547523$\cdot 10^{8}$ \\
  $\langle r^{8}_{+-} \rangle$ & 4.24772574985734547$\cdot 10^{9}$ & $\langle r^{8}_{--} \rangle$ & 
                                 8.66818739874051007$\cdot 10^{9}$ \\
             \hline
  $\langle r^{9}_{+-} \rangle$ & 1.514020468066805$\cdot 10^{11}$ & $\langle r^{9}_{--} \rangle$ & 
                                 3.072016630464095$\cdot 10^{11}$ \\
  $\langle r^{10}_{+-} \rangle$ & 5.99044264125978$\cdot 10^{12}$ & $\langle r^{10}_{--} \rangle$ & 
                                  1.21087163677764$\cdot 10^{13}$ \\

  $\langle r^{11}_{+-} \rangle$ & 2.6058671579139$\cdot 10^{14}$ & $\langle r^{11}_{--} \rangle$ & 
                                  5.2539948995694$\cdot 10^{14}$ \\
        \hline   
  $\tau_{31}$ & 0.5919817011489022332573754 & $\tau_{21}$ & 0.0197696328171320017563035 \\
  $\langle f \rangle$ & 0.0509332587787341170677527 & $\langle \frac{1}{r_{32} r_{31} r_{21}} \rangle$ & 
                                              0.02203423801633579310 \\
  $\langle \frac{1}{r_{32} r_{31}} \rangle$ & 0.090935346529989403556662 & $\langle \frac{1}{r_{31} r_{21}} \rangle$ & 
                                              0.060697690288581955139 \\
        \hline 
  $\nu_{+-}^{(a)}$ & -0.4999999999743 & $\nu_{--}$ & 0.49999999156 \\
                \hline 
  $\langle \frac12 {\bf p}^{2}_{1} \rangle$ & 0.0666192945358900085250295 & $\langle \frac12 {\bf p}^{2}_{3} \rangle$ & 
                                              0.1287664811612000907203387 \\

  $\langle {\bf p}_1 \cdot {\bf p}_2  \rangle$ & 4.4721079105799263297204503$\cdot 10^{-3}$ & $\langle {\bf p}_1 \cdot 
                            {\bf p}_3 \rangle$ & 0.12876648116120009072033872 \\

  $\langle {\bf r}_{31} \cdot {\bf r}_{21} \rangle$ & 46.5893169239906645003 & $\langle {\bf r}_{32} 
      \cdot {\bf r}_{31} \rangle$ & 1.82962030224729090960 \\
    \hline\hline
  \end{tabular}}
  \end{center}
${}^{(a)}$The predicted (or expected) electron-positron cusp equals -0.5 (exactly), while analogous electron-electron 
cusp equals 0.5 (exactly).
  \end{table}
\newpage
\begin{table}[tbp]
   \caption{Convergence of the expectation values of delta-functions determined for the ground $1^{1}S-$state of 
            the Ps$^{-}$ ion. The notation $N$ is the total number of basis functions used.}
     \begin{center}
%     \scalebox{0.80}{%
     \begin{tabular}{| c | c | c | c |}
      \hline\hline
 $N$  & $\langle \delta({\bf r}_{+-}) \rangle$ & $\langle \delta({\bf r}_{--}) \rangle$ & 
 $\langle \delta({\bf r}_{321}) \rangle$ \\ 
     \hline
 3400  & 2.07331980051456$\cdot 10^{-2}$  & 1.7099675635321$\cdot 10^{-4}$ & 3.038854$\cdot 10^{-5}$ \\ 

 3500  & 2.07331980051485$\cdot 10^{-2}$  & 1.7099675635144$\cdot 10^{-4}$ & 3.028396$\cdot 10^{-5}$ \\ 

 3600  & 2.07331980051509$\cdot 10^{-2}$  & 1.7099675634962$\cdot 10^{-4}$ & 3.002512$\cdot 10^{-5}$ \\ 

 3700  & 2.07331980051543$\cdot 10^{-2}$  & 1.7099675634923$\cdot 10^{-4}$ & 3.003051$\cdot 10^{-5}$ \\

 3800  & 2.073319800515179$\cdot 10^{-2}$ & 1.7099675634650$\cdot 10^{-4}$ & 3.049167$\cdot 10^{-5}$ \\

 3840  & 2.073319800515069$\cdot 10^{-2}$ & 1.7099675634846$\cdot 10^{-4}$ & 3.025447$\cdot 10^{-5}$ \\ 

 3842  & 2.073319800515057$\cdot 10^{-2}$ & 1.7099675634845$\cdot 10^{-4}$ & 3.025341$\cdot 10^{-5}$ \\ 
          \hline\hline
  \end{tabular}
  \end{center} 
  \end{table}
\begin{table}[tbp]
   \caption{The expectation values of a number of quasi-singular and singular properties (in atomic units) of 
            the ground (bound) $1^1S-$state in the Ps$^{-}$ ion. The notations $+$ and $-$ denote the positron 
            and electron, respectively.}
     \begin{center}
     \scalebox{0.95}{%
     \begin{tabular}{| c | c | c | c |}
      \hline\hline
  $\langle r^{-2}_{+-} \rangle$ & 0.2793265422249508 & $\langle r^{-2}_{--} \rangle$ & 0.0360220584545365 \\
        \hline
  $\frac12 \Bigl(\langle \frac{r^{2}_{32}}{r^{3}_{31}} \rangle - \langle \frac{r^{2}_{21}}{r^{3}_{31}} 
   \rangle\Bigr) $ & -0.1234320911052344965 & $\langle \frac{{\bf r}_{31} \cdot {\bf r}_{32}}{r^{3}_{31}} 
   \rangle$ & 0.0464784204243756567 \\
  $\langle \frac{{\bf r}_{31} \cdot {\bf r}_{21}}{r^{3}_{31}} \rangle$ & 0.29334260263484465 & 
  $\langle \frac{{\bf r}_{32} \cdot {\bf r}_{12}}{r^{3}_{21}} \rangle$ & 0.07781595282624420 \\
           \hline 
  $\langle r^{-3}_{+-} \rangle_{R}$ & -0.25348417470280099215 & $\langle r^{-3}_{--} \rangle_{R}$ & 0.011310500731864678 \\

  $\langle r^{-3}_{+-} \rangle$     & -0.17055138268219822395 & $\langle r^{-3}_{--} \rangle$ & 0.011994487757258498 \\
         \hline\hline
  \end{tabular}}
  \end{center} 
  \end{table}
\newpage
\begin{table}[tbp]
   \caption{Annihilation rates $\Gamma_{n\gamma}$ in $sec^{-1}$ determined for the ground $1^{1}S-$state 
            of the Ps$^{-}$ ion. The notation $n$ stands for the number of photons emitted during annihilation.}
     \begin{center}
%     \scalebox{0.80}{%
     \begin{tabular}{| c | c | c | c |}
      \hline\hline
  $\Gamma_{2\gamma}$ & $\Gamma_{3\gamma}$ & $\Gamma_{4\gamma}$ & $\Gamma_{5\gamma}$ \\  
      \hline
  2.08004810195$\cdot 10^{9}$ & 5.63523069413$\cdot 10^{6}$ & 3.0750689272$\cdot 10^{3}$ & 5.381655332 \\ 
      \hline\hline
  $\Gamma$ & $\Gamma^{(a)}$ & $\Gamma_{1\gamma}$ & --------  \\  
      \hline
  2.08568641313$\cdot 10^{9}$ & 2.08568333265$\cdot 10^{9}$ & 3.2223$\cdot 10^{-2}$ & ---------- \\
          \hline\hline
  \end{tabular}
  \end{center} 
 ${}^{(a)}$The total annihilation rate is determined from the formula, Eq.(\ref{Gammatot}).  
  \end{table}

\begin{thebibliography}{99}
\bibitem{zeta} To determine different integrals which describe various properties of the electron-positron 
gases one needs to know the following approximate numerical values of Riemann $\zeta-$function: $\zeta(5)$ = 
1.0369277551433699263314, $\zeta(7)$ = 1.0083492773819228268398, $\zeta(9)$ = 1.0020083928260822144179 and 
$\zeta(11)$ = 1.0004941886041194645587. For $n \ge 6$ we can assume (to very good accuracy) that $\zeta(2 n 
+ 1) \approx 1$.
\end{thebibliography}
\end{document}